# Influence of Polymers on the Nucleation of Calcium Silicate Hydrates

Andreas Picker [a], Luc Nicoleau [b], André Nonat [c], Christophe Labbez [c], Helmut Cölfen [a]*

[a] Physical Chemistry, University of Konstanz, Universitätsstraße 10, 78457 Konstanz, Germany

[b] BASF Construction Chemicals GmbH, GMB/M, 83308 Trostberg, Germany

[c] Université de Bourgogne, Faculté des Sciences Mirande, 21078 Dijon cedex, France

* e-mail: helmut.coelfen@uni-konstanz.de

## Abstract

This work focuses on the investigation of ongoing processes in the pre- and postnucleation stage of C-S-H precipitation at pH 12 and pH 13. Calcium induces the condensation of smaller silicate species to bigger oligomers in the prenucleation stage. By titration in combination with ion-selective electrodes, the effects of additives on the formation of C-S-H can be monitored and quantified in terms of calcium binding, nucleation times, supersaturation, and the post-nucleation behavior showing detailed differences between polymers. Negatively charged polymers inhibit nucleation, neutral or almost neutral polymers do not have an influence on nucleation time and supersaturation and the rare promotion of nucleation has been identified with cationic polymers. The post-nucleation data show the polymer influence on solubility and C/S ratio of the precipitated phase and stabilization of primary nucleated particles against further aggregation or growth. This work can be regarded as a step towards better control of C-S-H precipitation.

## 1. Introduction

Calcium silicate hydrate (C-S-H[i]) is known to be the most abundant hydration phase in hardened ordinary Portland cement pastes and thus mainly responsible for its cohesion and the final mechanical properties of cementitious materials [1, 2]. Because of its importance for the mechanical properties of cement-based construction materials, a better understanding of the phenomena underlying the cohesion, and then the possibility to further enhance this later, is highly desired. Yet, due to its poor crystallinity, a variable chemical composition and the extreme and complex conditions necessary for its formation in real systems (high pH, foreign ions, etc.), it has remained difficult to study this phase [3].

Although there is already quite some knowledge about mechanical properties of C-S-H, its stoichiometry or also the role of C-S-H in the cement paste there are still many open basic questions. For example it is still not fully understood, why the size of the platelet-like C-S-H crystallites [4, 5] is limited to the nanometer range [6] and most of all whether it can be modified. Moreover, the understanding of the growth of the C-S-H particles is still limited [7]. The preliminary step before growth is nucleation. As C-S-H crystals remain relatively tiny, it seems that secondary nucleation prevails over the crystal growth after a certain crystal size, making the nucleation even more important in the precipitation process. The C-S-H nucleation is poorly understood and is considered to be key in the development of the early mechanical properties. In a recent study by N. Krautwurst et al, the homogeneous nucleation of C-S-H was experimentally shown to proceed via a complex two-step pathway [8] resp. multi-step pathway [9] similar to that of calcium carbonate [10], Al(III) (oxy)(hydr)oxides [11], hematite [12] and other minerals [13]. This two-step nucleation pathway was also observed in the presence of superplasticizers [14, 15]. In the first step, stable amorphous spheroids are formed via an activated process, whose calcium and silicate content is diluted compared to the final crystal. In the second step these amorphous spheroids simultaneously aggregate and crystallize to tobermorite-type C-S-H. Still, the mechanisms responsible for this complex nucleation pathway are unclear and whether some precursor species (Ca-silicate complexes or clusters) are present in supersaturated solutions from the start, before the formation of the amorphous spheroids, remained an open question until the very recent publication of Sowoidnich et al. [9]. By means of analytical ultracentrifugation, they could show a number of species involved in the multi-step nucleation process of C-S-H. First silica dimers and oligomers with bound $Ca^{2+}$ are observed, which form prenucleation clusters with two different sizes. These form liquid droplets for which three different sizes were observed. They form an amorphous species with low density, which further densifies. These amorphous species are consistent with the amorphous spheroids observed by Krautwurst [8].

---
[i] In cement chemistry notation, C = CaO, S = $SiO_2$, and H = $H_2O$

Much effort has already been devoted to influence the cement hydration and workability of the cement paste with additives to tune the porosity and mechanical properties of the final material, but only if the ongoing processes at the early stage are understood, namely the nucleation/growth and the formation of hydrate network and C-S-H in particular, can those properties be modified in a controlled manner, see e.g. [16]. Besides the few published works mentioned above, there is, so far, not much research done yet in this field.

Here, we present a method that enables the quantification of effects of additives on the *in-situ* nucleation of C-S-H at various conditions. We adapted a titration assay which has already previously been reported for the investigation of nucleation processes of calcium carbonate [10, 17] as well as for the effects of polymers [18, 19], peptides [20] and amino acids [21] on its precipitation processes. However, the C-S-H system is significantly more complex. This is on the one hand due to its variable composition (the calcium to silicon ratio of C-S-H is dependent on the lime concentration in the surrounding equilibrium solution and thus also the pH), on the other hand due to the presence of silicates. Silicates were for example proven to influence the crystallization behavior of calcium carbonate in form of shifting the nucleation point towards higher supersaturations with respect to $CaCO_3$ [22]. Despite these difficulties, it was possible with the help of titrations and ion-selective electrodes to quantify the impact of different families of polymeric additives on the various steps of C-S-H formation.

For the investigation of the polymers on C-S-H nucleation, two sets of given conditions were chosen, which mainly differ in the pH value. On the one hand, pH 12 was chosen as "low pH" model system, while on the other hand, pH 13 was investigated as these conditions better reflect the environment in real cementitious systems. Although the experimental conditions have to be considered to be model conditions, the main goal of investigating the influence of polymeric additives on the in-situ formation of C-S-H could be addressed.

## 2. Experimental section

### 2.1. Materials

The following chemicals were all used as received: sodium metasilicate (Aldrich), calcium chloride (Fluka, 1M volumetric solution), sodium chloride (Riedel de Haën), sodium hydroxide (Merck, 1M volumetric solution), polyethylene glycol (PEG 8000, Aldrich), polyvinyl pyrrolidone (PVP 360000, Aldrich), polyvinyl alcohol (PVA 9000-10000, Aldrich), polystyrene sulfonate (PSS 70000, Aldrich), Goldschmidt EA3007 (PEG-b-PMAA, PEO 3000 g/mol, PMAA 700 g/mol, Evonik), polyacrylic acid (PAA 5100, PAA 100000, PAA 450000, Aldrich), poly(1-vinylpyrrolidone-co-acrylic acid) (PVP-co-PAA 96000, Aldrich), poly(styrene sulfonate-co-maleic acid) (PSS-co-PMA 20000, Aldrich), poly(acrylamide-co-acrylic acid) partial sodium salt (PAAm-co-PAA 200000 (≈20 wt.% acrylamide, ≈80 wt.% acrylic acid), PAAm-co-PAA 520000 (≈80 wt.% acrylamide, ≈20 wt.% acrylic acid), Aldrich). polyDADMAC with different molecular weights (28000 g/mol, 165000 g/mol and 941000 g/mol) was provided by the Frauenhofer Institute for Applied Polymer Research and Goldschmidt Phosphonated (GS-P, PEG-b-PMAA-PO$_3$H$_2$) with a phosphonation degree of 21% (determined by $^{31}$P-NMR) was synthesized from Goldschmidt EA3007. Chemical structures of the polymers are shown in Figure S1.

### 2.2. In-situ potentiometric titration measurements

Titration experiments were performed using a commercial, computer-controlled system from Metrohm (Filderstadt, Germany), operated with the custom-designed software Tiamo (v2.2). The setup consists of a titration device (Titrando 809) that regulates two dosing units (Dosino 807) capable of dispensing titrant solution in steps as small as 0.2 µl. The pH and free Ca$^{2+}$ concentration in the samples during titration were monitored in real-time utilizing a glass microelectrode (Metrohm, No. 6.0234.100) and a polymer-based ion-selective electrode (Metrohm, No. 6.0508.110), respectively. Electrodes, beaker and burette tips were cleaned with 0.1 M NaOH, distilled water, 10% acetic acid and again rinsed several times with distilled water after each experiment.

All experiments were performed at 23 ± 1°C. During all titrations, a gentle stream of water saturated argon was applied on the solution to avoid extensive uptake of atmospheric $CO_2$. In a typical run, 31.5ml (22.5 ml) of 3.7 mM $Na_2SiO_3$ (freshly prepared with degassed water in MilliQ quality) and 3.5 ml (2.5 ml) of polymer stock solution (1 g/l, water in case of the reference) were filled into a home-built sealed and argon flushed teflon beaker and adjusted to pH 12 (pH 13) by adding 0.55 ml (5 ml) of 1 M NaOH. 30 mM $CaCl_2$ was then added with a dosing speed of 0.02 ml/min while stirring at 700 rpm. Calcium potential and pH were monitored at the same time during the entire experiment.

The titration conditions differ from those of dissolution precipitation of cementitious particles in a way that in the titration experiment, the calcium ions are titrated into a silicate-rich solution in the beginning of the experiment with a fixed rate without a feedback loop to the chemical composition of the solution.

The ion-selective electrode was calibrated by titrating 30 mM $CaCl_2$ into water/NaOH pH 12 with additional amount of 4.3 mM NaCl (same ionic strength as 3.3 mM $Na_2SiO_3$) to have identical ionic strengths in the calibrations and the experiments. For pH 13, calibrations were done in water/NaOH pH 13.

For the thorough investigation of a titration curve in presence of PAAm-co-PAA 200000, 25 ml of PAAm-co-PAA 200000 stock solution (1 g/l) were added to 225 ml of 3.7 mM $Na_2SiO_3$ in a 500 ml beaker. The pH was adjusted to 13 by adding 50 ml of 1 M NaOH. Then, 30 mM $CaCl_2$ were titrated to the solution with a dosing speed of 0.05 ml/min while the calcium potential and pH were monitored.

Calcium binding experiments at pH 12 were performed to elucidate the adsorption capacity of $Ca^{2+}$- ions on the additives. For this, 3.5 ml polymer stock solution (1 g/l) was added to 31.5 ml $H_2O$, pH adjusted to pH 12 with 0.55 ml NaOH 1 M and 30 mM $CaCl_2$ was dosed while pH and calcium potential were monitored.

### 2.3. Mass spectrometry

ESI-MS measurements were carried out by direct infusion of the samples at a flow rate of 600 µl/h into an Esquire 3000+ ion trap mass spectrometer (Bruker Daltonik, Bremen, Germany) operated in positive ion mode. Mass spectra were recorded by scanning from 50 to 3000 m/z. The ion source parameters were as follows: 15 psi nebulizing gas (nitrogen), 3 l/min of drying gas (nitrogen) at a temperature of 25°C, capillary voltage 1500 V and capillary exit 104.5 V.

### 2.4. Adsorption measurements via total organic carbon analysis (TOC)

For adsorption measurements via TOC analysis, titration was normally stopped in the postnucleation stage after the addition of 4 ml (2.2 ml for pH 13) of 30 mM $CaCl_2$ to have equal amounts of ions and thus comparable quantities of precipitate and equal supersaturations with respect to C-S-H so that the results for both pH values are comparable. In special cases like PAA 100000 g/mol or PAA 450000 g/mol, more $CaCl_2$ was added to force nucleation. Dispersions were allowed to ripen for at least three days to ensure equilibrium conditions. After filtration, 3 ml of the filtrate were diluted with 16.6 ml $H_2O$ and acidified with 0.4 ml concentrated $H_3PO_4$. Additionally, reference samples with pure polymer were prepared. The amount of the measured carbon in the reference was correlated with the measured amount in the sample. The percentage of adsorbed polymer on C-S-H could then be obtained by subtracting the measured amount of polymer from the utilized one.

### 2.5. Dynamic light scattering (DLS)

For DLS measurements, samples were taken at different points of time of a C-S-H nucleation experiment and correlation curves were measured for one minute with an ALV-DLS. Measurements were performed in an infinite loop until the next sample was taken. The DLS-setup is using a Flex02-08D multiple tau digital correlator equipped with a ALV-SP 86 Goniometer (ALV, Langen, Germany) and a He-Ne laser (JDSU, CA,

USA) with a wavelength of ⊐ = 632.8 nm (max. power: <35mW). Measurements were performed at 25°C and a constant scattering angle of 90°.

### 2.6. Analytical ultracentrifugation

For ultracentrifugation experiments, 10 mM sodium silicate solutions were prepared in different concentrated NaOH (HCl for pH 11) to give final pH values of 11, 12 and 13. The samples were spinned at 60000 rpm and data analyzed with Sedfit. The sedimentation velocity data were evaluated by fitting of up to four discrete species to the Lamm equation. The ultracentrifuge is an analytical ultracentrifuge Beckman XL-I (Beckman Coulter, Palo Alto, CA) using RAYLEIGH interference optics for the detection of sedimenting species.

## 3. Results and discussion

### 3.1. *Nucleation of C-S-H in the absence of polymers*

In this work, we studied the influence of polymeric additives on the nucleation of calcium silicate hydrate in a $Na_2SiO_3$/$CaCl_2$ model system at two different pH values. For the purpose of interpreting the effects of the polymers, the focus was first of all on the pure C-S-H reference system.

Figure 1a) shows typical titration curves for the *in-situ* formation of C-S-H with the help of the titration setup in a $Na_2SiO_3$/$CaCl_2$ model system at pH 12 (black) and pH 13 (grey). The added amount of calcium is indicated by the dashed line. During the experiment, $CaCl_2$ is constantly added to a solution of sodium silicate (pre-adjusted to a distinct pH), leading to a linear increase of the calcium potential curves with time. Once the critical supersaturation with respect to C-S-H is reached, nucleation occurs. This is indicated by the peak maximum, the so-called nucleation point. The subsequent drop can then be attributed to the consumption of calcium ions due to the rapid C-S-H formation and growth so that these ions are not free in solution and thus not detectable anymore by the ISE. The formation of C-S-H under these model system conditions was confirmed by powder XRD measurements of the precipitates (Figure S2).

Depending on the preset pH value of the starting solution, the nucleation point is shifted to earlier or later nucleation times. This can be attributed to the different supersaturations ß with respect to C-S-H, which is highly dependent on the pH value. The higher the pH the higher the according supersaturation if all other ion concentrations in solution stay constant. Here, ß is defined as

$$ß = \frac{K_{supersaturation}}{K_{solubility}}$$

$$\beta = \frac{(Ca^{2+})^x \cdot (H_4SiO_4) \cdot (OH^-)^{2x}}{(Ca^{2+})^x_{eq} \cdot (H_4SiO_4)_{eq} \cdot (OH^-)^{2x}_{eq}}$$

with $K_{supersaturation}$ being the ion activity product of the supersaturated solution and $K_{solubility}$ the solubility product of the corresponding phase which was assumed to be constant in the following calculations. x is the calcium to silicon ratio of C-S-H considered. Supersaturations were calculated with PHREEQC [23] and the equilibria in the database are the same as in reference [23]. In this set of reactions, the Ca-Si complexes in solution have not been considered since it is not possible to fully and accurately describe this part of the solution chemistry in the presence of polymers anyway. Thus, the calculated ß values for the curves in Figure 1a) after 3200 s (nucleation point for pH 13) are approximately ß = 12 for pH 12 but already ß = 23 for pH 13 [24].

Another feature of the titration curves is their slopes. At pH 13, the curves follow exactly the straight line of the dosed amount of calcium. This means that every added calcium ion is free in solution and thus detectable by the calcium-selective electrode. At pH 12, the slopes of the curves are flatter and they differ from the dashed dosage line, indicating that free calcium ions are taken out of the system, most likely due to complexation by silicate in solution.

However, one has to be very careful when talking about the slope of the calcium curves at these conditions, due to two main experimental side effects. First of all, the calcium-selective electrode shows cross-sensitivities to several ions, among them also sodium. According to the manufacturers manual, concentrations of 260 mM Na$^+$ cause an error of ≈10% [25]. As the sodium concentrations at pH 13 are ≈160 mM this effect has to be taken into account. To avoid the effect of cross sensitivities during nucleation experiments, the calibration of the electrode was done under the same conditions as the subsequent measurements in the absence of silicate but in water/NaOH pH 13. Thus, the amount of sodium ions is equal and constant in both experiments and can be neglected in the following.

The second constraint is the ionic strength, which can alter the activity of the analyte and thus causes differences in the resulting slope [26]. Addition of 3.3 mM Na$_2$SiO$_3$ solution (in comparison to the calibration) can change the shape of the curves. At pH 13, this effect can be neglected as the high pH value ensures a constant ionic strength. The addition of 3.3 mM Na$_2$SiO$_3$ is not influencing the U$_{Ca}$ (both solutions having values of I = 15.5 mmol/l [24]). The situation changes for measurements at pH 12. Here, the ionic strength is ≈36% higher in the working solution (3.3 mM Na$_2$SiO$_3$, pH 12, I = 16.2 mmol/l [14]) as in the calibration solution (H$_2$O/NaOH, pH 12, I = 11.9 mmol/l [24]). This difference results in an error in the slope when the calibration is not done under the appropriate conditions. Figure 1b) shows the effect of different calibration conditions on the outcome of a C-S-H titration curve. The grey curve was obtained with calibration data from pure water/NaOH pH 12, the black one from water/NaOH pH 12 with an additional amount of 4.3 mmol/l NaCl as background electrolyte to adjust the ionic strength to the same value as during the measurement.

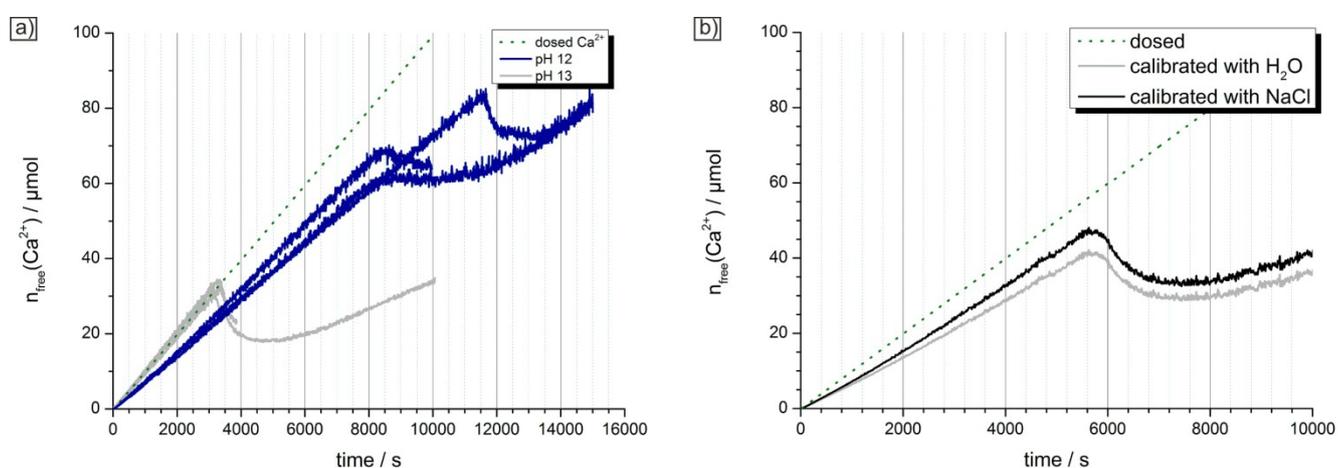

**Figure 1**. C-S-H titration curves. a) Comparison of pH 12 (black) and 13 (grey). [Na$_2$SiO$_3$] = 3 mM, [CaCl$_2$] = 30 mM, v = 0.02 ml/min. b) Influence of the ionic strength on the slope of the titration curve at pH 12. (1) grey: calibration in water/NaOH pH 12, (2) black: calibration in water/NaOH pH 12 with an additional amount of 4.3 mM NaCl for equal ionic strength conditions during calibration and experiment. The dashed line represents the amount of dosed calcium. [Na$_2$SiO$_3$] = 3.3 mM, [CaCl$_2$] = 30 mM, v = 0.02 mM.

Even if the slope of the measurement curve is increasing and approaching the dosed amount of calcium, the difference between added and measured ions remains. From this we can conclude that calcium is really bound in the prenucleation stage of C-S-H formation at pH 12, whereas no calcium is bound at pH 13. This can be explained by the different distribution of silica species under these conditions. From the calculated distribution profile (Figure S3, assuming only monomeric silicate in solution) it can be seen that the dominating species at pH 12 is the single negatively charged $Si(OH)_3O^-$, whereas its fraction decreases at higher pH and the double charged $Si(OH)_2O^{2-}$ gets more abundant. Thus, the increased pH is creating negative charges on the silicate ions, which results in electrostatic repulsion of the molecules in solution. Consequently, a silicate solution at pH 13 should be more stabilized against oligomerization and thus contain more silicate monomers and smaller chains than at pH 12. This assumption is experimentally supported by mass spectrometrical analysis (Figure S4) and ultracentrifugation results (Figure S5) showing both a trend towards smaller species with increasing pH which is moreover in good agreement with the literature [27-29].

Further on-line mass spectrometrical analysis was performed to shed light on the processes in a sodium silicate solution at pH 13 during titration of calcium (Figures S6 and S7). Here, it can be qualitatively observed that bigger species are developing with the ongoing addition of calcium until they vanish again towards the nucleation point, which can be attributed to an insufficient ionization so that the oligomers cannot be detected anymore by mass spectrometry. This outcome suggests a calcium mediated condensation of silicate species to higher oligomers in the prenucleation stage finally leading to the formation of C-S-H.

From the pH variation experiments in combination with analytical ultracentrifugation and mass spectrometry, it could be concluded that the pH has a high impact on the species distribution of the starting silicate solution, which is further supported by literature [27-29]. Oligomerization leading to bigger silicate species occurs for pH values of 11 and 12 whereas it is prevented or at least slowed down for pH 13. From titration experiments at different pH values, different degrees of calcium binding on silicate were observed, leading to the assumption that silicate monomers and smaller aggregates are not able to significantly bind calcium at low concentrations [30]. From mass spectrometical analysis an increasing amount of bigger species can be observed upon the incorporation of calcium into the system. These bigger particles vanish when the nucleation point is approached. From this it can be concluded that calcium induces the condensation of smaller silicate species to bigger oligomers as a first step towards the final formation of C-S-H. Convergent results were very recently obtained by the group of Fernandez-Martinez who used total scattering methods to characterize the structure of those oligomers seen as the building unit of C-S-H [31].

### *3.2. Nucleation of C-S-H in the presence of polymers*

After the investigation of the nucleation of C-S-H, the influence of polymers on the prenucleation regime of C-S-H will be addressed in the following by means of titration experiments at pH 12 and pH 13.

The titration curves in the presence of polymers are investigated regarding four main characteristics.[19] These are:

1. the time of nucleation (i.e. the peak maximum)
2. the supersaturation ß at the nucleation point with respect to C-S-H compared to ß = 23.7 and 19.5 for the reference cases at pH 12 and 13 (indicating an influence of the polymer on the silicate condensation and the species concentration necessary for the formation of the first stable nuclei) [24]
3. the amount of free calcium remaining in the solution (indicating the solubility of the precipitated phase and different Ca/Si ratios within the precipitate, respectively)
4. the amount of bound $Ca^{2+}$ by the polymer. Yet another quantitative determination of the $Ca^{2+}$ binding capacity values was done in reference experiments at pH 12 in absence of silicates. Data are presented in Table 1.

These characteristics are illustrated schematically in Figure 2, where the possible influences of the polymers on the titration curves are highlighted and compared to a titration curve recorded in the absence of an additive (black curve).

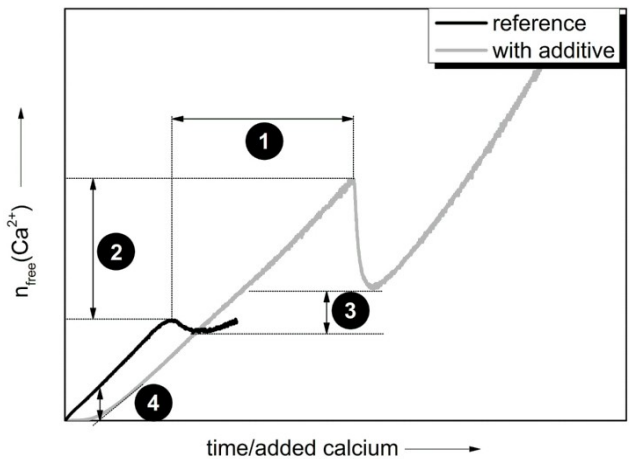

**Figure 2.** Qualitative development of the free amount of $Ca^{2+}$ in solution with time and addition of $CaCl_2$ into a $Na_2SiO_3$ solution, respectively. The influence of polymers on the nucleation of C-S-H was described by the following four criteria: (1) the time of nucleation (showing possible delaying or accelerating effects of polymers), (2) the supersaturation at the nucleation point ß (indicating the influence of polymers on the energy required to form first stable nuclei), (3) the level of free calcium after nucleation (indicating the amount of precipitated calcium and thus the Ca/Si ratio in the solid C-S-H and the solubility, respectively), (4) $Ca^{2+}$-binding of polymers. Thus, the hypothetical additive depicted in the graph does: (1) delay nucleation, (2) favor the nucleation at higher supersaturations with respect to C-S-H and thus stabilize the solution against nuclei formation, (3) favor the precipitation of a more soluble phase or with higher Ca/Si ratio and (4) bind $Ca^{2+}$. The slope of the post-nucleation curve contains information on post-nucleation $Ca^{2+}$ binding. If the slope is the same as that of the added $Ca^{2+}$, no $Ca^{2+}$ binding takes place after nucleation anymore.

In real cementitious systems, the control of the nucleation time and thus the supersaturation during the prenucleation regime can have huge impacts on the mortars and/or concrete properties. Indeed, the precipitation of first hydrates will lead to (1) the stiffening of the mix indicating the end of the workability period and (2) the start of the setting or in other words the start of the very early mechanical properties enabling sometimes the handling of concrete elements. The mechanical strength of cementitious binders depends amongst others on the volume fraction filled by emerging hydrates during the hydration of $C_3S$ [32, 33]. One possible benefit of nucleation inhibition by polymers is directly linked with the supersaturation level in solution. Delaying the nucleation and formation of C-S-H may allow for more $C_3S$ to be dissolved at the very early stage and finally result in greater amounts of C-S-H and thus in the development of higher mechanical strength. This is indeed, what is generally observed with a number of additives (acting as retarders) and specific range of dosages, where the initial delay is more than simply compensated by the acceleration period of the cement hydration. That is, significantly higher hydration degrees with additives than without are observed at a later stage. However, the exact mechanisms responsible for this behavior are still not very well understood, in part because most of these additives are also known to slow down the dissolution of the cement clinker [34].

Besides nucleation retarders, accelerating agents are beneficial as well. As the setting and hardening is highly dependent on the formation of C-S-H, these processes could be accelerated if the nucleation barrier of C-S-H nucleation is energetically decreased. Like this, fast setting concretes could be realized, which gets particularly important in modern processes in which faster and greener construction is requested. This increase of reactivity can also mean a decrease of the energy spent during the curing of precast elements, and thus a lower amount of $CO_2$ embedded in the final concrete. The acceleration of cement hydration by salts like NaCl and $CaCl_2$ is already reported in the literature [4]. It could be shown, that the number of initially precipitated C-S-H nuclei in a suspension of $C_3S$ is increased in the presence of NaCl, which finally speeds up the hydration process of $C_3S$. Moreover, ion-exchange resins were reported to accelerate the early-age hydration [35].

For the purpose of quantifying the nucleation time compared to the reference cases, we introduce here a retarding factor $F_R$ defined as:

$$F_R = \frac{\text{average nucleation time} \in \text{presence of additives}}{\text{average nucleation time} \in \text{absence of additives}}$$

Figure 3 shows exemplarily some curves obtained from titrations in the presence of polymers. The complete data are presented in the SI. Starting from these curves, the data discussed in the following were extracted from the average peak maxima (nucleation time), the concentrations of calcium measured by the electrode and silicate (calculated, all dilutions taken into account) at the peak maxima (determination of ß) and the concentration of free calcium remaining after nucleation in solution (impact on the calcium content in the precipitate, all dilutions taken into account). For the determination of the calcium binding capability of the polymers, calcium solution was dosed into a polymer solution at pH 12 while pH and calcium potential was recorded. As the ion selective electrode only measures free ions in solution, the difference of dosed and measured calcium can be attributed to the binding of calcium as it is exemplarily shown in Figure S49. The complete data including $F_R$-, ß- and $Ca^{2+}$ binding values are summarized in Table S4.

Additionally, all polymers were investigated regarding their adsorption capability on C-S-H after the precipitation in titration experiments. For this, titrations were stopped after reaching equal supersaturations with respect to C-S-H for both pH values to obtain comparable results. The adsorption level for both pH values are shown in Tables S1 and S2 (all dilutions taken into account). Surprisingly, all polymers show very high adsorption ratios over 80%. This can probably be ascribed to the very little amount of additive that was added during the nucleation experiments (0.1 g/l), resulting in unspecific binding because all polymer molecules are able to find a binding site. Going to higher polymer concentrations, one would expect rather different adsorption rates that are more dependent on the chemical functions of the polymers. However, the outcome of these experiments has the advantage that efficient adsorption on C-S-H as an additional variable can be neglected so that all observed variations of the titration curves can be attributed to the polymer structure itself in the following.

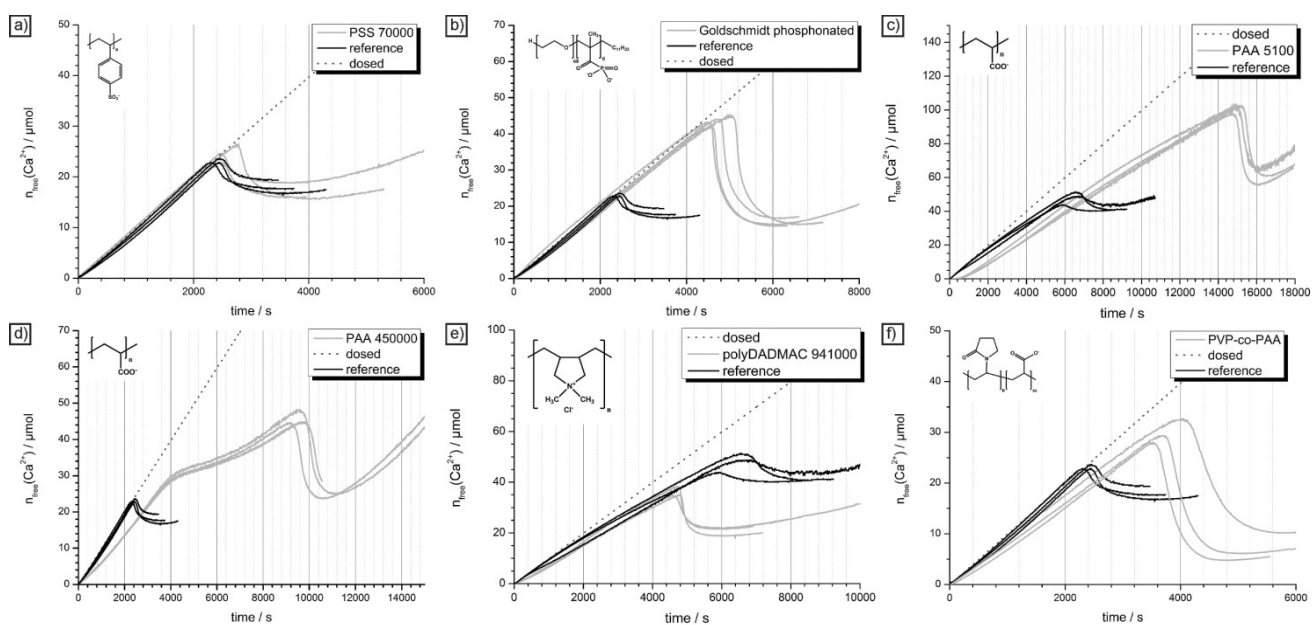

**Figure 3.** Examples of titration curves that show the discussed features. Experiments in presence of additives (grey curves) are compared to reference experiments in absence of additives (black curves). c) and e) for pH 12, others for pH 13. a) PSS 70000, no influence, b) Goldschmidt phosphonated, nucleation retardation, c) PAA 5100, nucleation retardation and $Ca^{2+}$ binding, d) PAA 450000, nucleation retardation, $Ca^{2+}$ binding and stabilization of primary nucleated particles, e) polyDADMAC 941000, nucleation acceleration and precipitation of a calcium rich phase and f) PVP-co-PAA, precipitation of a calcium rich C-S-H.

For the following discussion of the influence of polymers on the nucleation of C-S-H, the additives are divided into five groups which are: (1) neutral polymers, (2) negatively charged polymers with high charge density, (3) block co-polymers, (4) statistical co-polymers and (5) positively charged polymers with high, permanent charge density.

### *3.2.1. Time of nucleation and corresponding supersaturation at the nucleation point*

Figure 4a) and 4d) illustrate the influence of polymers on the average nucleation time at pH 12 and pH 13, Figure 4b) and 4e) the corresponding supersaturation values ß at the nucleation point. Measurements at pH 13 turned out to be much more reproducible with much less fluctuations than at pH 12. This might be attributed to the ten times higher ionic strength caused by NaOH, which is acting as a background electrolyte and thus stabilized the potential measured by the ion selective electrode.

The polymers can be categorized into three groups according to their mode of action: additives that (1) have neither influence on the nucleation time and nor on the calcium concentration, i.e, nor on ß, (2) show nucleation retardation towards higher supersaturations and (3) show nucleation acceleration towards lower supersaturations.

3.2.1.1. Polymers without an effect on nucleation time and calcium concentration

Among the polymers without any effect at both pH values are PEG (Figures S8 and S27), PVP (Figures S9 and S28), PSS (Figures S10 and S30) and PVA (Figure S29). For PEG, PVP and PVA, this outcome can be rationalized by their charge neutrality and thus their inability to interact with prenucleation species. However, similar behavior is observed for PSS. Here it was already shown elsewhere that PSS (Mw=70000 g/mol) was able to alter the shape and morphology of calcium carbonate crystals and even form $CaCO_3$ mesocrystals [36, 37]. Although the additive is powerful in this respect, it could be shown that the effect of PSS in the prenucleation stage of $CaCO_3$ formation is limited as in the present study [19]. This outcome for both systems is rather surprising as PSS is a strong polyelectrolyte and bears negative charges due to the deprotonated sulfonate groups, which could adsorb on C-S-H or $CaCO_3$. One explanation could be the delocalization of the negative charge over three oxygen atoms, making the interaction with $Ca^{2+}$ less attractive and/or making the competition with $Na^+$ a bit more favorable for $Na^+$. Indeed, in $Ca^{2+}$ binding experiments at pH 12, it could be shown that PSS is able to bind around 0.13 $Ca^{2+}$ ions per monomer (Table 1). At pH 9.75, this value was determined to be 0.31 ions per monomer [19]. The difference could be explained by the 100 times higher $Na^+$ concentration at pH 12, which competes with the $Ca^{2+}$ ions for free binding sites on the polymer.

**Table 1.** Summary of the effects of polymeric additives on the titration curves of C-S-H at various conditions. For the calculations of the $Ca^{2+}$ binding capability of negatively charged polymers at pH 12 only charged monomers are taken into account, e.g. only PAA for PVP-co-PAA. Functional side groups: C = carboxylate, S = sulfonate, P = phosphonate. The bound $Ca^{2+}$ per monomer unit was only calculated for pH 12, since at pH 13, almost no binding occurred (with the exception of PAA) as also reflected in the much smaller missing $Ca^{2+}$ at the nucleation point.

| Polymer | Functional side group | $M_W$ [g/mol] | $F_R$ at pH 12 | SD | $F_R$ at pH 13 | SD | ß at pH 12 | SD | ß at pH 13 | SD | Missing $Ca^{2+}$ at nucleation pH 12 | SD | Missing $Ca^{2+}$ at nucleation pH 13 | SD | Bound $Ca^{2+}$ at pH 12 per monomer |
|---|---|---|---|---|---|---|---|---|---|---|---|---|---|---|---|
| Ref. | --- | 0 | 1 |  | 1 |  | 24,5 | 1,64 | 19,5 | 0,45 | 15,1 | 1,08 | 0,5 | 0,56 | --- |
| PSS | S | 70000 | 0,9 | 0,02 | 1,07 | 0,07 | 22,4 |  | 21,4 | 1,48 | 10,9 | 1,47 | 0,9 | 0,4 | 0,13 |
| PVP | --- | 360000 | 1,32 | 0,15 | 1,04 | 0,04 | 28,3 | 3,71 | 19,8 | 1,99 | 20,2 | 1,86 | 1,4 | 0,85 | --- |
| PEG | --- | 8000 | 0,93 | 0,01 | 1,1 | 0,05 | 23,3 | 1,73 | 20,3 | 2,7 | 14,4 | 1,4 | 1,6 | 0,75 | --- |
| PVA | --- | 9000 | --- |  | 1,28 | 0,15 | --- |  | 26,1 | 4,15 |  |  | 0,9 | 0,1 | --- |
| GS-P | P, C | 4000 | 1,52 | 0,14 | 1,96 | 0,10 | 29,3 | 1,1 | 38,3 | 1,12 | 34,1 | 6,61 | 3,2 | 1,4 | 0,07 |
| GS | C | 3950 | 1,32 | 0,08 | 1,21 | 0,06 | 33,4 | 3,2 | 23,7 | 1,655 | 16,5 | 5,06 | 1,4 | 0,81 | 0,03 |
| PAA | C | 5100 | 2,32 | 0,04 | 1.65 2.87 | 0,00 | 45,8 | 1,08 | 28.2 35.5 | 2,87 | 47,8 | 1,39 | 7,3 | 1,78 | 0,18 |
| PAA | C | 100000 | 2,75 | 0,07 | 1.66 4.42 | 0,02 | 50,8 | 4,79 | 19.6 31.1 | 3,07 | 64,2 | 3,55 | 15,3 | 2,5 | 0,53 |
| PAA | C | 450000 | --- |  | 1.66 3.97 |  | --- |  | 25.1 36.4 | 0,34 |  |  | 10,9 | 0,62 | 0,34 |
| PVP-co-PAA | C | 96000 | 1,51 | 0,02 | 1,58 | 0,11 | 39,4 | 5,33 | 26,2 | 2,22 | 16,7 | 6,03 | 7,3 | 0,19 | 0,18 |
| PSS-co-PMA | S, C | 20000 | 1,93 | 0,11 | 2,5 | 0,12 | 49,7 | 4,04 | 38,4 |  | 15,9 | 6,9 | 13,7 | 2,25 | 0,51 |
| PAAm-co-PAA | C | 200000 | 2,09 | 0,07 | 1.99 3.06 | 0,08 | 38 | 2,96 | 31,7 | 2,91 | 49,5 | 1,52 | 11,5 | 2,22 | 0,27 |
| PAAm-co-PAA | C | 520000 | 1,3 | 0,05 | 1,04 | 0,02 | 32,3 | 1,8 | 20,3 | 0,71 | 16,3 | 1,5 | 0,6 | 0,9 | 0,05 |
| polyDADMAC | --- | 28000 | 0,73 | 0,06 | 0,96 | 0,03 | 15,9 | 0,95 | 18,2 | 1,34 | 13,6 | 2,41 | 1,6 | 1,27 | --- |
| polyDADMAC | --- | 165000 | 0,68 | 0,01 | 0,91 | 0,01 | 15,2 | 1,75 | 17,7 | 0,2 | 12,9 | 2,2 | 1,1 | 0,34 | --- |
| polyDADMAC | --- | 941000 | 0,73 | 0,01 | 0,96 | 0,02 | 16,2 | 0,9 | 19,1 | 0,78 | 11,1 | 1,31 | 1,5 | 0,66 | --- |

3.2.1.2. Nucleation inhibiting polymers

The retarders are in principle all negatively charged polymers with carboxylic functions including the (partly) charged co-polymers and also GS-P with a partly phosphonated function. If the $Ca^{2+}$ binding to the several functional groups on the polymer is correlated to the solubility of the respective minerals reflecting the lattice energy, the following binding affinity of the polymers to $Ca^{2+}$ can be qualitatively deduced: Phosphates / phosphonates > carbonates > sulfates / sulfonates.

However, the highest efficiency in nucleation inhibition is observed for the pure acrylic acids PAA 5100 (Figures S13 and S33), PAA 100000 (Figures S14 and S34) and PAA 450000 (Figure S35) as well as for PSS-co-PMA (Figures S16 and 37) and PAAm-co-PAA 200000 having an acrylic acid content of ≈ wt. 80% (Figures S17 and 38). These polymers show additionally the highest negative charge density resulting also in the highest $Ca^{2+}$ binding affinity of all investigated polymers at pH 12 (ranging from 0.18 – 0.53 $Ca^{2+}$ ions per charged monomer, see Table 1). The two features of nucleation inhibition and $Ca^{2+}$ binding seem to correlate (Table 1, Figures S13 -S17 and Figures S33 – S38). The nucleation points at pH 12 are retarded by factors up to 2.75 (PAA 100000, Table S4). As the polymers bind calcium ions at the same time, an explanation for this finding might simply be the reduced amount of free ions within the solution [38, 39]. To refute this claim, the time being necessary to dose exactly the initially adsorbed amount of calcium is determined and illustrated exemplarily for PAA 5100 in Figure S 49. PAA 5100 is able to bind 0.18 $Ca^{2+}$ ions per acrylic acid monomer at pH 12 which corresponds to 240 $\mu$mol/l for a 0.1 g/l polymer solution. This amount is reached in the nucleation experiments after 850 seconds. The nucleation delay, however, is 8070 seconds, so nearly by a factor of 9 higher. Thus, the decrease in the amount of free calcium *via* complexation by PAA 5100 cannot be responsible for the observed delay [38].

As the chain length of the poly(acrylic acid) seems to play a role in the inhibition of C-S-H nucleation, a comparison of PAA 5100 and PAA 100000 is shown in Figure S22. The first observation is the higher calcium complexing capability of PAA 100000 than PAA 5100 even if the concentration (0.1 g/l) and thus the number of acrylic acid monomers is the same. This is also reflected in the quantitative measurements of the $Ca^{2+}$ binding capability. PAA 5100 is, as mentioned above, able to bind 0.18 $Ca^{2+}$ ions per monomer at pH 12, for PAA 100000 this value is already 0.53 while it is decreasing again to 0.34 for PAA 450000 (see Table 1). Interestingly, this agrees with the ability of PAA to form hydrogels by $Ca^{2+}$ crosslinking as "mineral plastic" precursors, where PAA 50000 - 100000 leads to the optimum gel with equal solid (G') and liquid (G") part of the frequency dependent shear modulus, which was thermodynamically explained by a complex balance between molar mass dependent $Ca^{2+}$ binding enthalpies as well as entropic reasons due to ion hydration water release and loss in PAA conformational entropy by $Ca^{2+}$ crosslinking [40].

The overall trend of molar mass-dependent ion binding has also been confirmed by Monte Carlo simulations (Figure S48). Even if the absolute values differ from the simulated ones, the trend towards higher adsorption capabilities of longer chains is given. This suggests that the secondary structuring of polymers can make a difference. The assumption of conformational changes is supported by results from Porus et al. [41]. With the help of optical reflectometry and quartz microbalance measurements, they found that polyelectrolytes adsorbed on a surface change their conformation from a rather flat conformation at low salt concentrations to a more globular one when electrolyte concentrations are increased [41]. However, this cannot explain the decreased $Ca^{2+}$ adsorption for PAA 450000 as compared to PAA 100000. Here, it is likely that such a long polymer precipitates upon calcium addition so that the binding capability is decreased again [42].

The second interesting finding about the chain length comparison in Figure S22 is the slightly higher retarding capability and ß value of PAA 100000 (50.8 *vs.* 45.8 for PAA 5100, Table S4). This correlates well with the increased calcium affinity. Comparing the C/S ratio of the precipitated C-S-H, this value is not significantly different compared to the reference, despite the much higher supersaturation in the system. The slopes of the polymer curves are nearly parallel to the theoretical dosing straight-line in the postnucleation stage, indicating, that every additionally added calcium ion remains free in solution. As C-S-H nuclei should have been formed and calcium ions are available, the explanation for this non-consumption would be rather the lack of silicate ions.

Figure S23 shows the comparison of different concentrations of PAA 100000 on the nucleation of C-S-H at pH 12. It can be seen that the $Ca^{2+}$ binding is dramatically increased with increasing PAA concentration. Moreover, the nucleation of C-S-H is strongly retarded for 1 g/l (around 22000 s corresponding to $F_R$ = 3.40) and even more for 2 g/l (around 50000 s corresponding to $F_R$ = 7.73), indicated by the black arrows while ß is decreased to 16.3 and 16.6. The much flatter slope suggests that a lot of calcium is bound to silicates *via* the polymers. This behavior was suggested by Sowoidnich et al. who identified, by means of analytical ultracentrifugation, polymer-ion clusters with a size of ~4nm in the aqueous phase of $C_3S$ paste after 10 minutes of hydration in the presence of superplasticizers [14]. It was further confirmed very recently in the PhD work of L. Bouzouaid who used gluconate as a simple model for PCE superplasticizers [43]. A zoo of hetero polynuclear complexes involving organic molecules as well as calcium, hydroxide and silicate species could be identified from measurements and simulations of the C-S-H solubility. From the fact that only a slight peak can be observed in the titration curve of 1 g/l and 2 g/l PAA 100000 without displaying the typical subsequent drop in the calcium potential, we can conclude that the formed species are stabilized immediately after their formation against further aggregation and growth. This means in turn, that the polymers are acting in both the pre- and post-nucleation stage of C-S-H formation.

One special feature can be identified for the nucleation curves at pH 13 in the presence of negatively charged polymers containing carboxylate functions. This group contains exclusively polymers with high negative charge density in form of poly(acrylic acid) with different polymerization degrees (Figures S33 – S35), PSS-co-PMA (Figure S37) and also PAAm-co-PAA 200000 (80 wt.% acrylic acid, Figure S38). All of the investigated polymers show a flatter prenucleation slope in combination with a more or less pronounced plateau after a first nucleation point, before a belated drop in the calcium potential indicates particle growth or probably the formation of secondary nucleated particles *via* a heterogenous (secondary) nucleation process on primarily formed nuclei, which results in accelerated growth of C-S-H particles. The polymers stabilize the primary formed particles to a certain extent until the entire additive is consumed so that subsequent particles can nucleate heterogeneously or further growth occurs. This is actually the strategy used by the X-Seed® technology developed by BASF to accelerate the cement hydration at an early age [44].

To some extent, it may also explain the significantly faster increase and higher values of hydration degrees obtained during the acceleration period in the presence of superplasticizers. The plateau for PAAm-co-PAA 200000 is in this respect different from those of the poly(acrylic acids). It is much flatter and nearly horizontal which means that every additional added calcium ion is immediately utilized to stabilize primary particles while the implementation is slower in the case of PAA.

The comparison with the PAA batch reveals, that the first bending of the curve always occurs at about the same time, as reflected by the nearly indistinguishable $F^1_R$ values, ranging from 1.65 - 1.66 (Figure S 43 and Table S4). This suggests that the primary nucleation of C-S-H is not altered with the polymerization degree of PAA. Even if PAA is capable of retarding the nucleation of C-S-H, this feature is independent of the chain length. As stated before, the dominating species in the prenucleation regime of C-S-H nucleation are silicate oligomers. Those are apparently small enough to be adsorbed to PAA chains *via* $Ca^{2+}$ bridging regardless of the polymer length. However, when it comes to the nucleation of primary C-S-H particles, the magnitude of their stabilization is highly altered indicated by the corresponding $F^2_R$ values ranging from 3.06 for PAA 5100 to 3.97 for PAA 450000 and 4.42 for PAA 100000 (Table S4). These data show clearly the superior capability of long chain polymers over shorter ones to stabilize primary nucleated C-S-H particles. Interestingly, PAA 100000 is more efficient in this respect compared to PAA 450000, which correlates well with the obtained values from the calcium binding experiments, where PAA 100000 was shown to adsorb more calcium than PAA 450000 (0.53 *vs*. 0.34 $Ca^{2+}$ bound per monomer, Table 1). The comparison between the ß values reveals that the nucleation occurs always at equal supersaturations with respect to C-S-H. Even if the stabilization of primary formed particles is prolonged, secondary nucleation or growth always occurs at comparable supersaturations independently of the polymer chain length.

As the entire feature of plateau formation between the first and second nucleation point (see for example Figure S34 and S35) is only displayed by polymers that are able to bind calcium ions it might be assumed that these properties are correlated, at least to a certain extent.

Lastly, Figure S 44 shows another concentration series of PAA 100000 at pH 13, revealing the same features as at pH 12 (Figure S 23). While the variation of the polymer chain length did not show any effect on the nucleation inhibition, the concentration indeed does. It can be seen that the very little amount of only 0.01 g/l PAA 100000 (orange curve) has already the same (according to the $F^1_R$ value of 1.92 even a bigger) effect on nucleation inhibition as a ten times higher concentration. However, when it comes to the stabilization of primary formed particles, this little amount fails completely, so that the titration curve drops down without forming a plateau. Going to higher polymer concentrations of 1 g/l and 2 g/l, the nucleation point is further shifted to higher retarding values ($F^1_R$ of 5.49 and 10.19) but lower supersaturation (ß of 24.5 and 22.9). Higher nucleation retardation at simultaneously decreasing supersaturation indicates an additional pronounced calcium binding of the deprotonated carboxylate groups. For both high polymer cases, no secondary particle formation was observed even after 70000 s.

The Goldschmidt PEO-b-PMAA based block co-polymers GS (Figures S11 and S31) and GS-P (Figures S12 and S32) are much less efficient regarding both, nucleation inhibition and $Ca^{2+}$ binding compared to the previous polymers. A feasible explanation for this observation is the shorter chain length both bearing only 8 charged monomers per molecule. The two remaining co-polymers PVP-co-PAA (Figures S15 and S36) and PAAm-co-PAA 520000 (Figures S18 and S39) show similar effects in this respect as GS and GS-P. Consisting of the same monomers as PAAm-co-PAA 200000, PAAm-co-PAA 520000 with only ≈ 20 wt.% of acrylic acid underlines the assumption, that a high negative charge density is needed for efficient nucleation inhibition. The same is true for PVP-co-PAA (≈ 25 wt.% acrylic acid), delaying slightly more than PAAm-co-PAA 520000.

With the two Goldschmidt polymers only differing in the functional charged group (carboxylate for Goldschmidt and 21% phosphonate for Goldschmidt phosphonated, determined by $^{31}$P-NMR), they exhibit an excellent possibility to elucidate the effectiveness of these two different negatively charged residues regarding their influence on C-S-H nucleation. The curves for both additives at pH 12 are summarized for comparison reasons in Figures S24. Their retarding factors ($F_R$ = 1.32 *vs*. 1.52) and ß values (33.4 *vs.* 29.3) are quite similar, although the phosphonated polymer retards a bit more. However, it has to be taken into account that in the phosphonated block copolymer, only 1 -2 of the 8 methacrylic acid groups are phosphonated so that the difference between these block copolymers is small. As the supersaturations are rather similar, it can be assumed that the oligomerization of silicates is similar. However, the slope of the phosphonated case is flatter indicating more calcium binding by the silicates in the prenucleation stage due to the polymer. Otherwise, the curves are only differing in the solubility of the nucleated phase.

For pH 13, the comparison of GS and GS-P is shown in Figure S45. Although the titration curves of these two polymers were nearly identical in the previous experiments at pH 12, they reveal clear differences at pH 13. The phosphonated polymer seems to be much more efficient regarding nucleation inhibition than the carboxylated one. Moreover, the supersaturation at the nucleation point is highly increased (23.7 *vs.* 38.3). From studies on calcium carbonate it is known, that phosphonates bind much better to calcium than carboxylates do. However, it has to be taken into account that the phosphonated additive bears two negative charges per phosphonated monomer due to the rather low p$K$-values of phosphonic acid (p$K_1$ = 2.0, p$K_2$ = 6.6) [45], while the carboxylated has only one. Due to the double negative charge, calcium adsorption is more efficient even in presence of a high excess of sodium ions, which is also manifested in the slightly higher calcium binding efficiencies at pH 12 (Table 1).

As the Goldschmidt phosphonated polymer is quite effective regarding nucleation inhibition at pH 13, the concentration of this additive was increased to 1 g/l, as is shown in Figure S46. Besides a much higher $F_R$ value (3.08) compared to the standard polymer case ($F_R$ = 1.96) the titration curve now shows qualitatively the same features as PSS-co-PMA above. Towards the nucleation point it starts to bend indicating that more calcium is consumed as during the steeper region before. This phenomenon will be discussed in more detail in the following.

Since the silicate species and the polymers are negatively charged, calcium plays a role as bridging agent between both. It was concluded before that the oligomerization of silicates into larger chains [8] is an important step towards the formation of C-S-H. We speculate that these oligomers could be complexed by negatively charged polymers *via* Ca$^{2+}$ bridging and thus be stabilized against subsequent aggregation and nucleation.

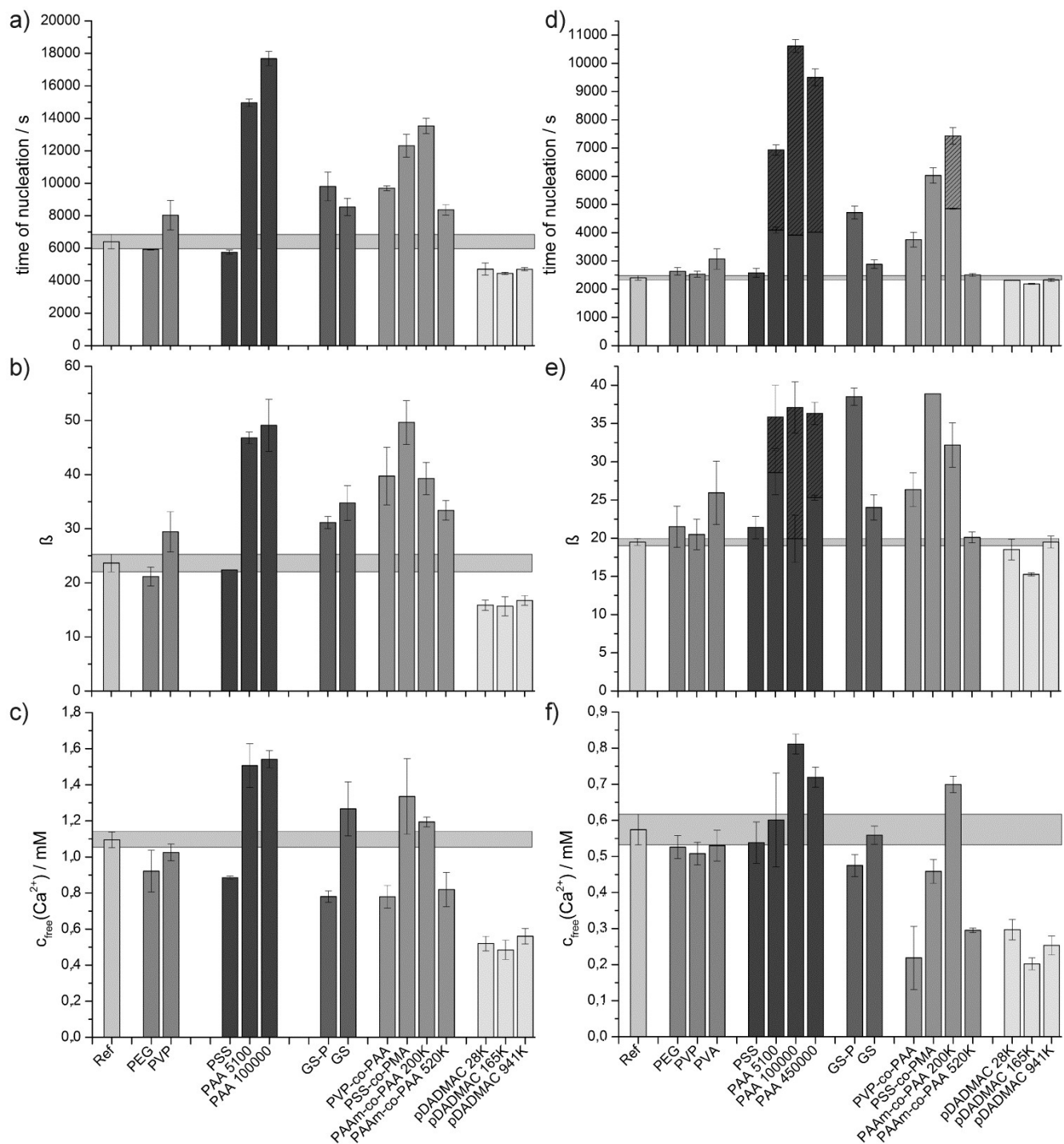

**Figure 4:** Bar plots illustrating the effect of polymers on the nucleation of C-S-H at pH 12 (left column) and pH 13 (right column) with respect to time of nucleation (first row), supersaturation ß at the nucleation point (second row) and concentration of free calcium remaining in solution after nucleation (third row). Bars are all shown with error bars representing ± one standard deviation. For comparison reasons, values from reference experiments in absence of polymers are shown and highlighted via the horizontal bar in light grey. Shaded parts of the bars in d) and e) represent the values for the formation of secondary nucleated particles.

3.2.1.3. Nucleation accelerators

A very interesting feature is displayed by the positively charged polyDADMAC polymers. All of them show independently of their chain length an acceleration effect on the nucleation of C-S-H at pH 12 (Figures S19 – S21), which is reflected in the calculated $F_R$ values ranging between 0.69 – 0.74 (Table S4). The behavior is very special as so-called *nucleators* are very rare. Indeed, for calcium carbonate being one of the most investigated and best understood compound regarding nucleation and crystal growth, no such additives have ever been found to date. However, in silica biominerals, polycationic peptides were found to catalyze silica polycondensation [46]. Similarly, polyamines as well as various amino acids were reported to catalyze the nucleation and growth of silica [47, 48]. Interestingly, Dove et al found that these organics mostly affect (here decrease) the kinetic barrier to the nucleation (kinetic pre-factor of the classical nucleation theory) but not the Gibbs free energy barrier [49]. In other terms, the organics facilitate the transfer of silicate species from the solution to the nucleus but do not decrease its surface free energy. They further observed that the decrease of the kinetic barrier is all the more pronounced as the organics are positively charged. This scenario observed for silica [49] and gypsum [50] may also be valid to explain the acceleration of C-S-H nucleation in presence of polyDADMAC. When first nuclei are formed, free silicate and free calcium can feed the growth of C-S-H, indicated by the drop of the calcium potential curve. The entire feature of nucleation acceleration of polyDADMAC at pH 12 was shown to be independent of the chain length (Figure S25).

Going to ten times higher polyDADMAC concentrations, the situation changes and there are now twice as much DADMAC monomers than silicate molecules. The acceleration effect is much less pronounced ($F_R$ = 0.92) (Figure S26). Similarly, PolyDADMAC at pH 13 (Figures S40 - S42) does not show the acceleration effect as dramatically as at pH 12 ($F_R$ ranging from 0.68-0.73, Table S4) anymore. Amongst all tested polymers they remain, however, the one with the lowest $F_R$ values (ranging from 0.91-0.96). This behavior may be understood as a consequence of the complexation of silicate by polyDAMAC, which counteracts its catalytic effect. At higher concentrations or pH (at pH 13 silicate is doubly charged), there is less free silicate in solution as more silicate is bound to polyDADMAC. Thus, the apparent supersaturation with respect to C-S-H is significantly reduced, which results in lower nucleation rates but also to C-S-H with higher calcium to silicon ratio as indicated by the lower amount of free calcium in solution. Other possible reasons are the precipitation of the polyDADMAC with $H_2SiO_4^{2-}$, higher affinity of $H_2SiO_4^{2-}$ to $Ca^{2+}$ than between $H_3SiO_4^{-}$ and $Ca^{2+}$, making the binding of $H_2SiO_4^{2-}$ with $DADMAC^+$ less attractive. Also, at pH 13, the hydroxide ion may also significantly compete with $H_2SiO_4^{2-}$ or $H_3SiO_4^{-}$ for being the counterion of $DADMAC^+$. Conversely, there is much more $Na^+$ which can compete with $DADMAC^+$ for complexing the silicates.

### *3.2.2. Free calcium after nucleation*

Figure 4c) and f) compare the polymers with respect to their impact on the free calcium measured after the nucleation. This parameter can be addressed to shed light on the calcium content of the solid.

At pH 12 (Figure 4c)) there are 7 polymers that significantly lower the free calcium concentration after the nucleation of C-S-H (PSS, GS-P, PVP-co-PAA, PAAm-co-PAA 520K and the polyDADMACs) and only 3 that increase it (PAA 5100, PAA 100000 and PSS-co-PMA). Looking at the PAAs, PVP-co-PAA 200K (80 wt.% PAA) and PAAm-co-PAA 520K (20 wt.% PAA) the trend for the negatively charged residues seems to be the same as for the times of nucleation. The more negative charge the higher is the remaining free calcium in solution and the lower the calcium content in the resulting precipitate. On the contrary, PVP-co-PAA and positively charged residues as the polyDADMACs favor the incorporation of more calcium into the resulting solid.

### 3.3. *Investigation of a Titration Curve in Presence of PAAm-co-PAA 200000*

In the previous part, the influence of different kinds of polymers was investigated by means of titration experiments at various pH values. One special feature could be observed for the poly(acrylic acids) and PAAm-co-PAA 200000 at pH 13, where a well pronounced plateau was found. As this behavior was regarded peculiar, more thorough investigations of these titration curves were performed and will be discussed in this section. For this purpose, the titration curve of PAAm-co-PAA 200000 was exemplarily taken for the entire group in which a plateau is observed at pH 13. pH was monitored during the nucleation experiment and on-line dynamic light scattering, WAXS and EDX were performed at different stages.

### 3.3.1. Titration curve analysis

First of all, the titration curve in combination with the pH value monitoring will be examined in Figure 5. The calcium potential curve displays the same behavior as already described before with a plateau forming between two nucleation points. Taking a look at the corresponding pH curve reveals that (1) the pH continuously decreases with the addition of calcium because of the formation of CaOH$^+$ species and (2) the pH is dropping with the first bending of the calcium potential curve. As the formation of C-S-H is according to Equation 1, this drop is certainly correlated to the OH$^-$ consuming formation of C-S-H nuclei (indicated by the black arrow in Figure 5).

$$xCa^{2+} + 2(x-1)OH^- + H_2SiO_4^{2-} \rightarrow (CaO)_x - (SiO_2) - (H_2O)_y$$

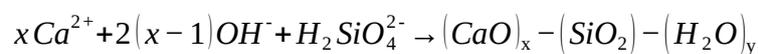

**Equation 1.** Formation of C-S-H in aqueous solution. The silicate species is formally written as H$_2$SiO$_4^{2-}$. This is of course only an approximation as the silicate speciation in aqueous solution is much more complicated. It is dependent on many factors, amongst them the concentration, ionic strength and pH. Moreover, the possibility of condensation reactions and hence the occurrence of diverse soluble oligomeric species has to be taken into account.

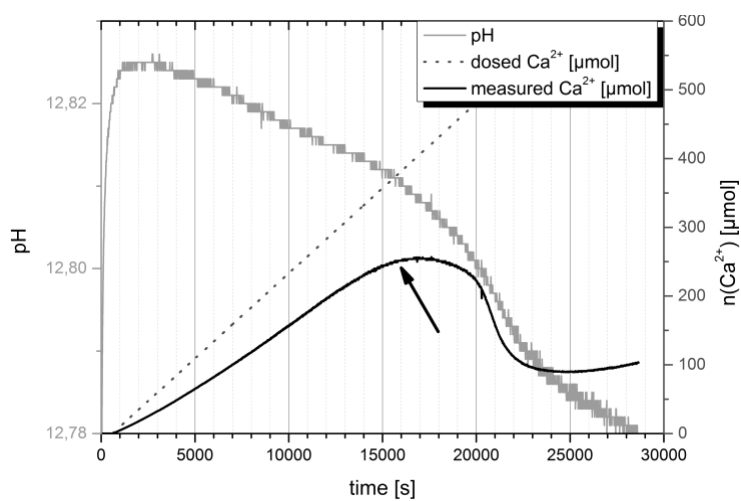

**Figure 5:** Calcium and pH curve of C-S-H at pH 13 in presence of PAAm-co-PAA 200000 ([Na$_2$SiO$_3$] = 3.7 mM, [CaCl$_2$] = 30 mM, v = 0.05 ml/min). The black arrow indicates the first nucleation point.

### 3.3.2. On-line dynamic light scattering

To confirm the presence of nucleated particles already at this early stage of the titration curve, on-line dynamic light scattering experiments were performed during C-S-H nucleation experiments. Samples were taken after different volumes of added calcium and measured in an endless loop until the next sample was taken from the solution. In this way, a rather smooth light scattering curve could be obtained and correlated with the titration outcome. Both particle size and scattering intensity were recorded and plotted together with the calcium potential curve, as shown in Figure 6. It is evident that no detectable species (> 10 nm) are present in solution at the beginning of the experiment. Only after around 12000 s, first particles appear, which is even before the drop of the pH value. However, their concentration is rather low, indicated by the intensity curve keeping a rather constant value until 13000 s. Afterwards both the intensity and the detected particle size start to give clear information. Particles with radii of 140 ±40 nm can be detected and also the intensity curve indicates an increasing concentration of nucleated species. Regarding the calculated particle size, it has to be noted that the fits of the auto correlation functions assume spherical particles. Therefore, the quantitative analysis of the DLS sizes is not accurate, as C-S-H rather precipitates in form of plates or fibrils but not spheres. Moreover, depending on the polymer, the plates form more or less pronounced aggregates in solution, finally resulting in the observed radius of around 140 nm.

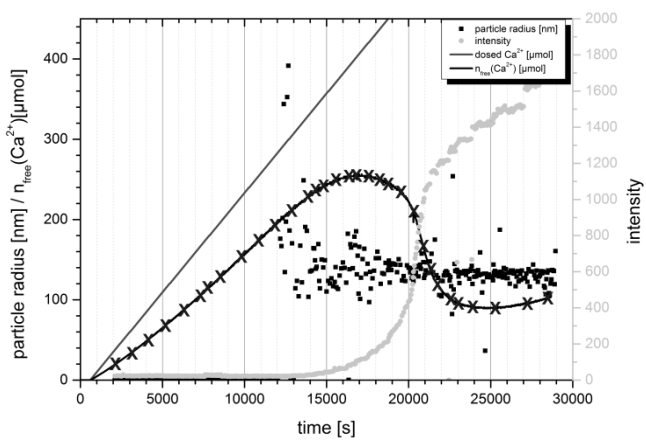

**Figure 6:** Online DLS measurements of the titration curve of C-S-H at pH 13 in presence of PAAm-co-PAA 200000. Crosses on the nucleation curve indicate the time points where samples were taken ([Na$_2$SiO$_3$] = 3.7 mM, [CaCl$_2$] = 30 mM, v = 0.05 ml/min).

When it comes to the strong bending of the curve and the beginning of the plateau around 15000 s, respectively, the signal becomes more intense, which causes also the particle size to get much more stable and monodisperse due to the better fitting quality. However, it is only at the second nucleation point that the intensity signal is increasing exponentially and flattens again when the calcium potential curve reaches its minimum, to finally end up in a sigmoidal shape. The particle size remains constant during this intensity development, until the stabilized C-S-H particles start to aggregate to a micrometer-sized precipitate after some days. Thus, particles are still stabilized in the post-nucleation regime by the polymer even after the precipitation. PAAm-co-PAA seems to prevent aggregation of nucleated C-S-H to a certain extent, probably due *to* an increased steric repulsion of the particles.

From the pH and the intensity curve progression as well as from the particle size development, we can assume that first stable particles nucleate homogeneously from solution already with the beginning of the bending of the calcium potential curve rather than only at the peak maximum. From the concentration variation experiments for PAA 100000 (Figure S44) and the scattering intensity in Figure 6, we can conclude that during the plateau more and more nuclei are formed (intensity increase) and stabilized (concentration variation) until the polymer is consumed. Reaching this point, a further process has to occur that consumes a lot of calcium and $OH^-$. Due to an increasing amount of nucleation sites devoid of polymers is formed, accelerated C-S-H particle formation occurs.

After the drop in the calcium potential, the scattering intensity increases further but much slower until a value of around 2000 at 80000 s while the particle size stays constant. This indicates that further C-S-H is formed from solution until the corresponding calcium and silicate equilibrium concentrations are reached.

### 3.3.3. *X-ray diffraction analysis*

To gain insight in the type of formed particles at various points of time, X-ray diffraction experiments were performed for the three stages of the titration curve: One sample in the prenucleation regime, one in the plateau and the last one after the drop of the calcium potential. As particles are formed in the last two stages, they could simply be separated with 200 nm filters, washed with water and EtOH and dried at 40°C.

Yet, it is more difficult in the prenucleation regime to collect the formed species. Therefore, the entire solution was poured into a 10 times excess volume of EtOH to quench the solution and to force the nucleation, because the species are less soluble in EtOH. Then, the milky dispersion was filtered with 50 nm filters due to the smaller species than in the case of precipitates and washed once with EtOH. Due to the isolation method, special care has to be taken regarding the possible artifacts and thus the conclusions.

The spectra of the 3 samples are shown in Figure 7. Here, it can be observed that the final product (curve 3) exhibits the characteristic broad peaks of C-S-H at values of $2\Theta$ = 29°, 32°, 50° and 55°, which confirms the formation of C-S-H. However, this peak pattern develops stepwise. Starting from an amorphous broad peak for the prenucleation phase (curve 1) an enhancement of crystallinity is observed during the plateau step and finally in the post-nucleation step. The most intense peak at 29° is clearly visible, whereas the 32° peak can only be identified as a shoulder. The same is true for the two other signals at 50° and 55°, the latter being completely missing and not identifiable, respectively, due to the low signal to noise ratio.

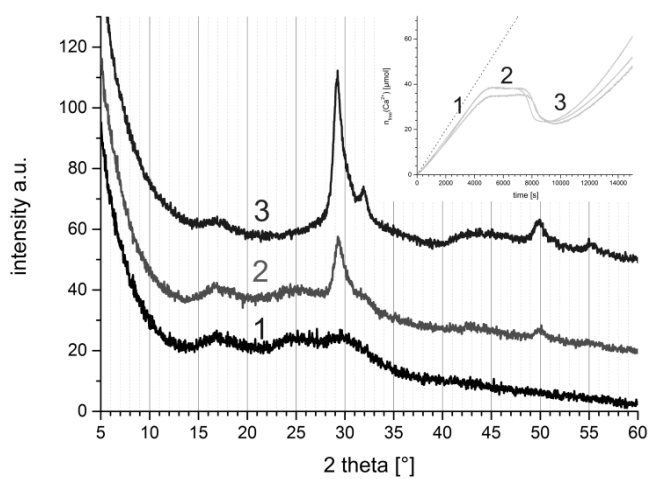

**Figure 7:** Powder XRD measurements at different points of time of the titration curve at pH 13 (small inlet) in presence of 0.1 g/l PAAm-co-PAA 200000. The inset shows the titration curve with the points where the samples were taken.

From the X-ray diffraction experiments, we can conclude that a non or low-crystallinity material dominates in the precipitate collected during the prenucleation regime. EDX analysis of the isolated prenucleation species reveals that it contains all elements present in solution, i.e. silicon, calcium and sodium (Figure S51).

The second XRD spectrum in Figure 7 corresponds to the formation of first particles and the flattened calcium potential curve, respectively. Here, the bulk material starts to exhibit ordering without displaying the final crystallinity of C-S-H. One feasible scenario for these results is based on the previous mechanism proposed in the first part but has to be slightly modified due to the observed plateau-like transition state because each added $Ca^{2+}$ is consumed. Such a formation process in presence of PAAm-co-PAA could imply the following steps:

1. calcium-catalyzed oligomerization of silicates to higher oligomers
2. prestructuring of silicate chains due to calcium bridging of higher oligomers [51]
3. stabilization of rather unordered primary nucleated particles by the polymer
4. a) precipitation of secondly nucleated particles showing higher structural ordering due to a higher C/S ratio in solution caused by the decreased silicate concentration

    b) once ordering and crystallization occurs, further growth proceeds

5. stabilization of small particles, i.e., inhibition of aggregation by the polymer

Most likely, the different proposed steps are the same as for the non-additive case, with the difference that the first nucleated particles are stabilized by the polymer whereas they agglomerate immediately in the absence of stabilizing agents. While primary nucleated particles do not show a fully pronounced crystallinity yet but only prestructuring, it is developed over the course of time in the additive-free case and after the second nucleation in the presence of PAAm-co-PAA.

## 4. Conclusions

In this work, we could show that the applied method of titration in combination with ion selective electrodes is a powerful method for the investigation of nucleation processes, even in case of complex systems like calcium silicate hydrate. Due to its simplicity, rapidity and effectiveness titration, especially with ion sensitive electrodes, is a good method for the quantitative evaluation of the polymer impact on the formation of solid compounds.

As it has been recently highlighted the oligomerization process is key in the formation of first C-S-H nuclei. The oligomerization strongly depends on the deprotonation state of silicate species and thus on pH. Oligomerization leading to bigger silicate species occurs for pH values of 11 and 12, whereas it is prevented or at least slowed down for pH 13. It has been shown that calcium binding on silicate species also depends on pH, leading to the assumption that silicate monomers and smaller aggregates are not able to significantly bind calcium.

Yet at pH 13, an increasing amount of bigger species can be observed upon the incorporation of calcium into the system. These bigger particles vanish when the nucleation point is approached. From this it can be concluded that calcium induces the condensation of smaller silicate species to bigger oligomers in the prenucleation stage of the C-S-H formation.

The polymers show different effects according to their chemistry. **Negatively charged polymers retard the nucleation,** i.e., a higher supersaturation is required to nucleate C-S-H. The retardation is not directly linked to the affinity to calcium. Indeed, some polymers retard the nucleation without binding a lot of calcium ions. It can be suggested that such a polymer either interacts with the nucleating species through other ions, or, simply prevents the oligomerization process via steric hindrance [50] without being strongly bound to the (pre-)nucleation species. The effect of polyDADMAC suggests that **positively charged polymers promote the nucleation**. At this stage without further information, it can be only proposed that cationic polymers either kinetically support the oligomerization because of a higher concentration of silicate species close to cationic functions, or, reduce the energy necessary to the formation of first nuclei because they interact with their surface and decrease the crystal-solution energy. **Neutral or almost neutral polymers (PEG and PVP) do not significantly influence** the C-S-H nucleation under these conditions. Another peculiar behavior has been identified. Some polymers like **PAAm-co-PAA 200000** lead to a calcium concentration plateau despite the calcium addition. Further analysis showed that **primary nucleated particles are stabilized** with pre-oriented but not yet fully pronounced C-S-H structure.

## Acknowledgements:

The authors acknowledge BASF SE for financial support of this project.

**Supporting Information**

### 5. pH dependent analysis

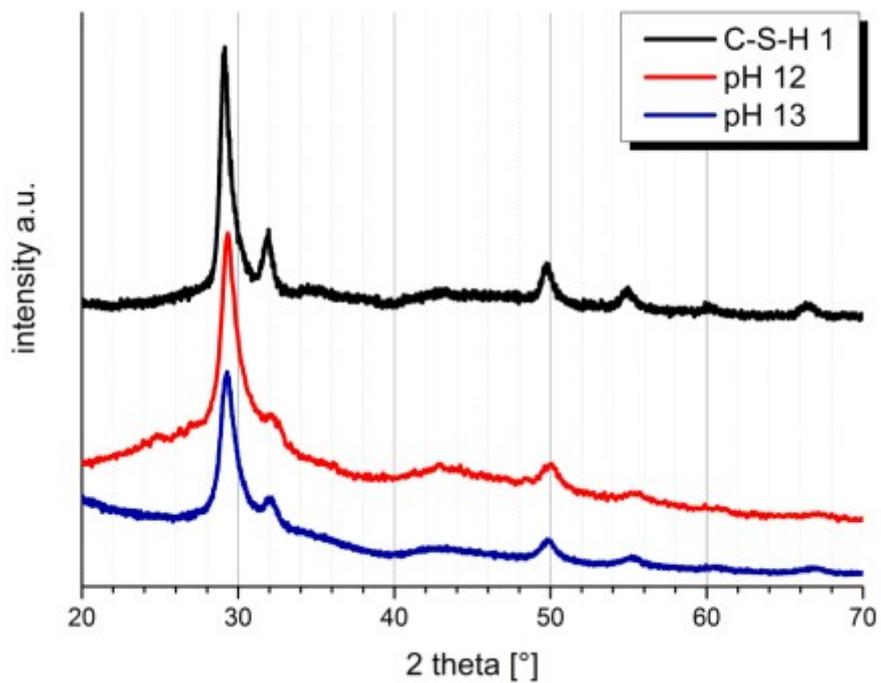

**Figure S 1.** Powder XRD pattern confirming the precipitation of C-S-H in titration experiments at pH 12 and pH 13 in $CaCl_2/Na_2SiO_3$ model systems. For comparison reasons, a pattern of pure C-S-H 1 (C/S ratio = 1) from the pozzolanic synthesis (CaO + $SiO_2$) is given.

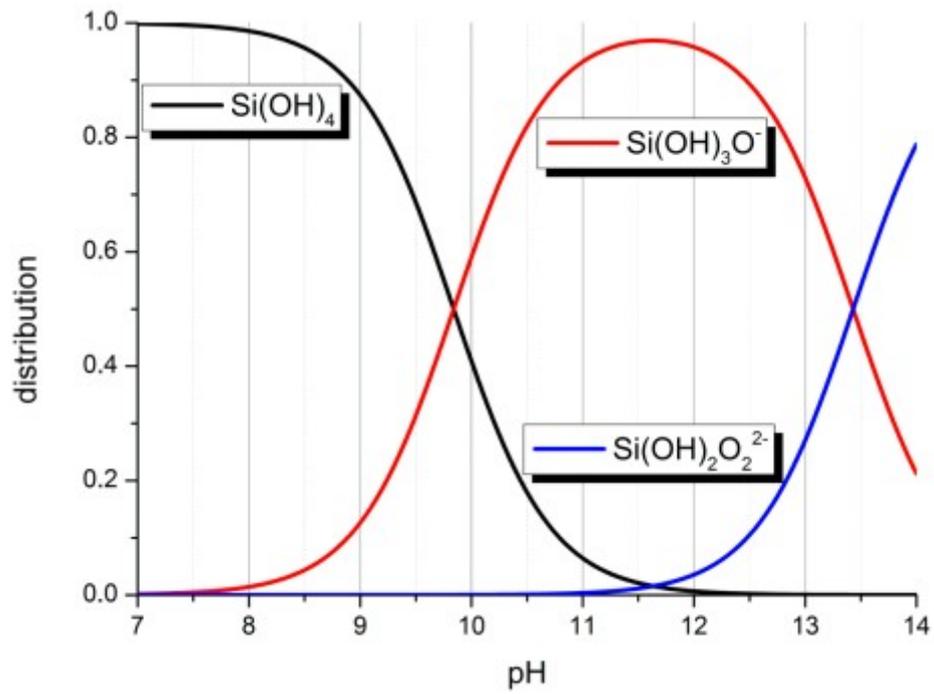

**Figure S 2.** Calculated distribution profile of monomeric silica species assuming only monomers and infinite dilution. Equilibrium constants are $K_1 = 1.44*10^{-10}$ and $K_2 = 3.72 * 10^{-14}$, taken from Dove and Rimstidt 1994.

## 6. Polymers

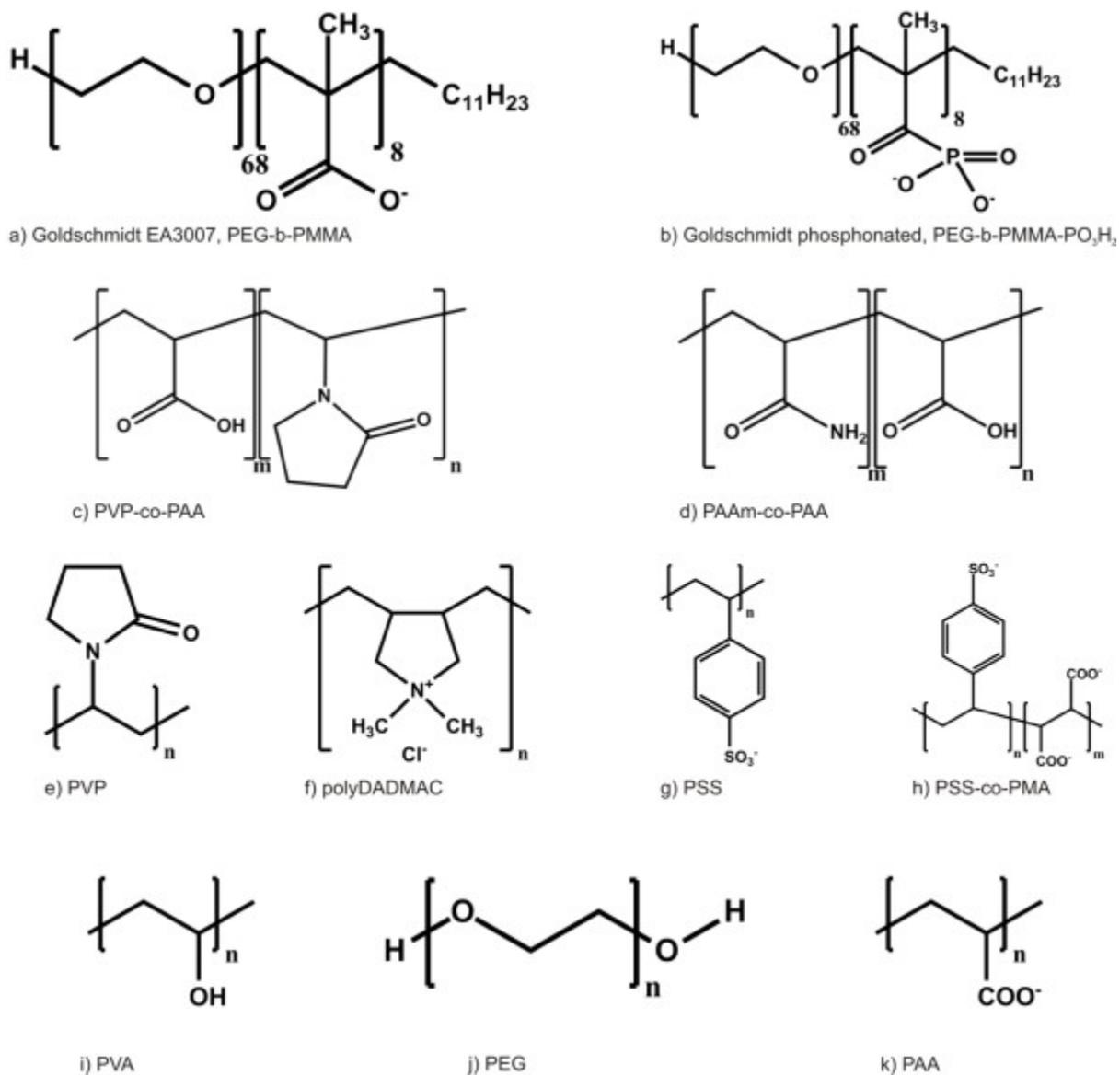

Figure S 3. Chemical structures of polymers utilized during the titration experiments.

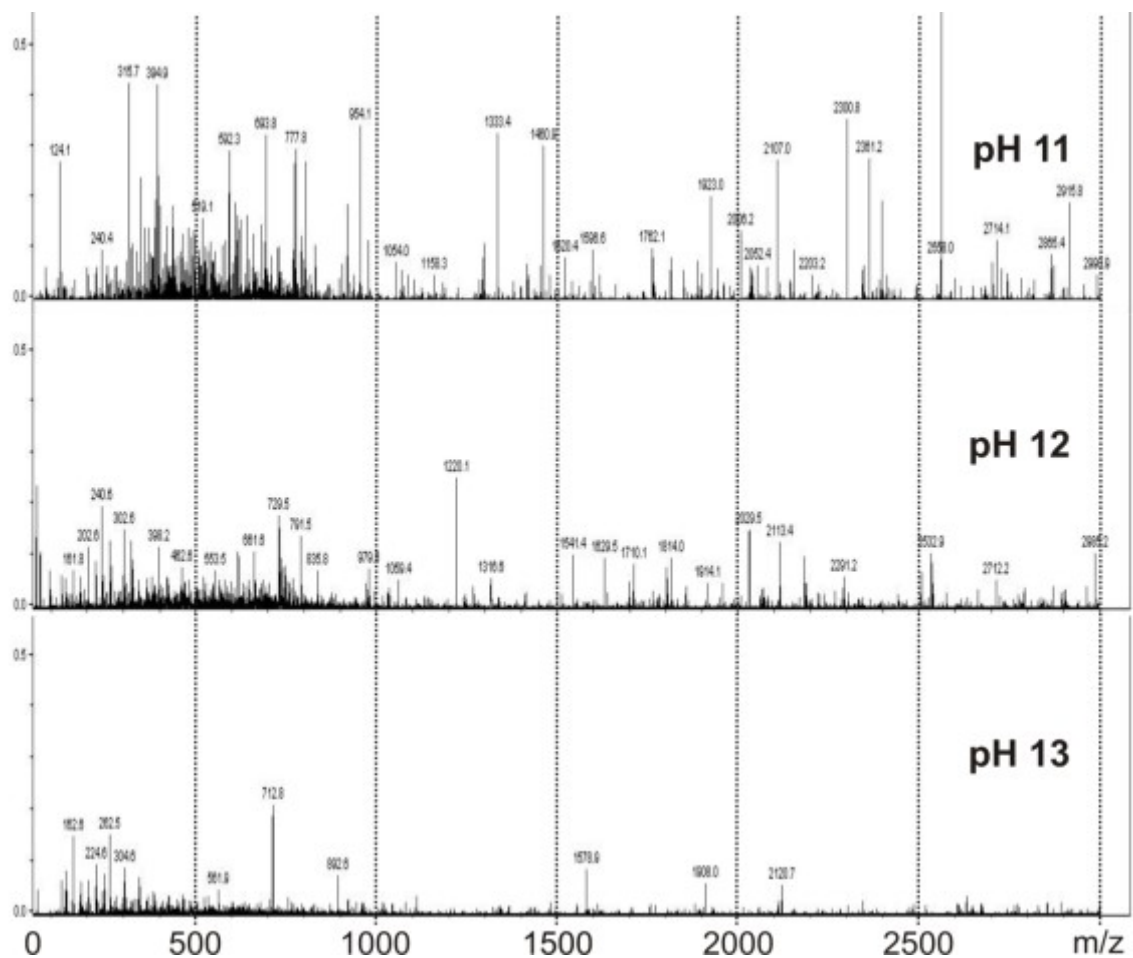

**Figure S 4.** Mass spectrometrical analysis of 3.3 mM sodium silicate solutions at pH 11, 12 and 13. Bigger species vanish with increasing pH value indicating that oligomer formation of silicates is hindered due to electrostatic repulsion.

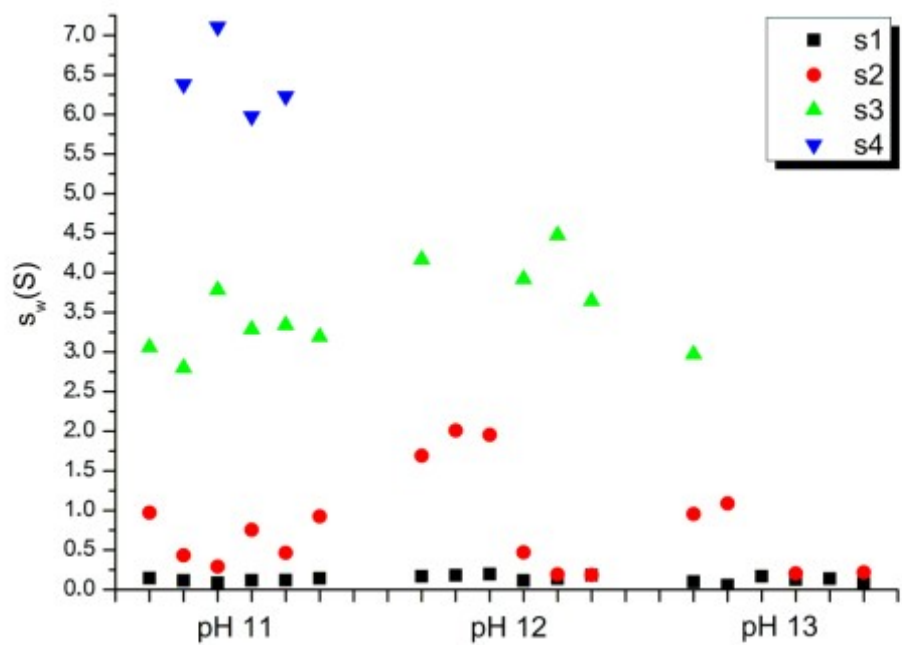

**Figure S 5.** Weight average sedimentation coefficients at 25°C identified *via* analytical ultracentrifugation analysis of 10 mM sodium silicate solutions at pH 11, 12 and 13. The different colors indicate different identified species. The black squares with sedimentation coefficients up to 0.5 S can be referred to ions and small molecules. The additional species indicated by blue and green triangles as well as the red circles correspond to larger ones. In the case of silicate, these additional species can be attributed to oligomers of different sizes. However, it has to be taken into account that there are no distinct, sharp sedimentation coefficient distributions obtainable from silicate samples as oligomers are polydisperse in solution. Nevertheless, the trend towards smaller species with increasing pH is given.

7. **Mass spectrometrical analysis of the reference titration curve at pH 13**

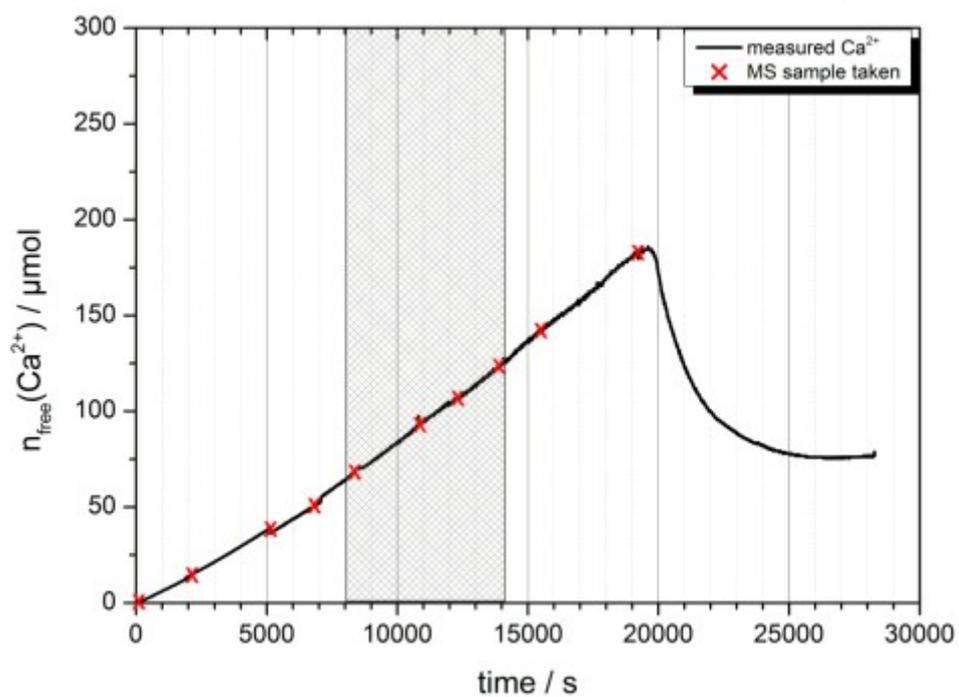

**Figure S 6.** Titration curve of C-S-H at pH 13 followed with mass spectrometry. Red crosses indicate the points of time when samples were taken. The region, where higher oligomers are detected is shaded in grey. Parameters: [$Na_2SiO_3$] = 3.3 mM, [$CaCl_2$] = 30 mM, v = 0.02 ml/min.

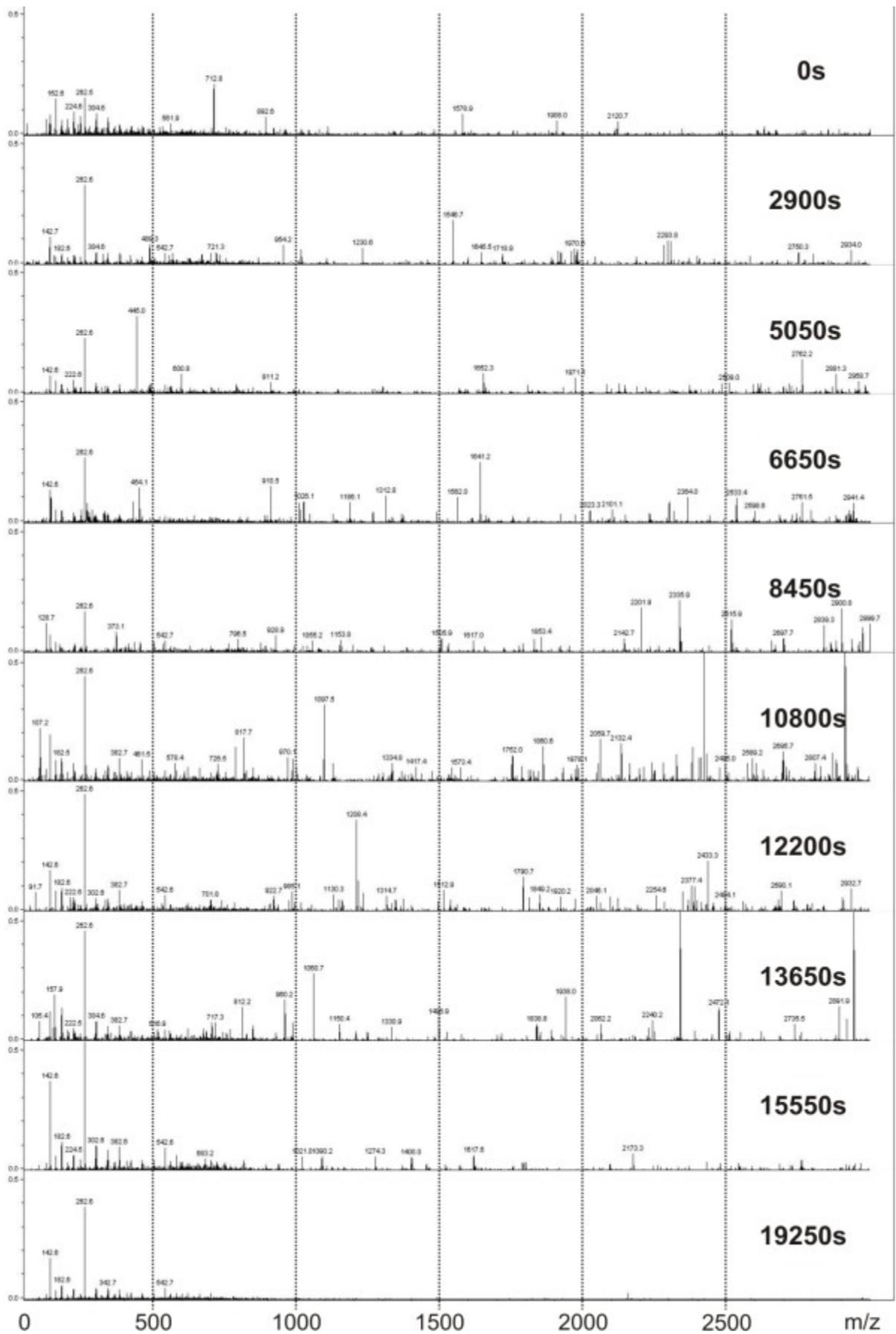

**Figure S 7.** Obtained mass spectra of samples from Figure S5 at various points of time during the C-S-H titration at pH 13. At the beginning of the experiment, predominantly smaller species are observed with m/z values up to 1000, which is due to the electrostatic repulsion of silicates at high pH stabilizing them against further oligomerization. After starting the addition of calcium, peaks between 1000 and 3000 m/z begin to appear, reaching their maximum amount between 10800 and 13650 s. Upon further incorporation of calcium ions into the system, the amount of larger species decreases until they vanish nearly completely at 19250 s, which is just before the nucleation point indicated by the calcium electrode.

## 8. Adsorption of polymers on C-S-H at pH 12

| Polymer | $V_{added}(CaCl_2)$ [ml] | Adsorbed on C-S-H [%] |
|---|---|---|
| PEG | 4.00 | 92.5 |
| PVP | 4.00 | 82.2 |
| PVA | 4.00 | 83.0 |
| PSS | 4.00 | 94.9 |
| GS-P | 4.00 | 86.1 |
| GS | 4.00 | 87.2 |
| PAA 5100 | 4.00 | 97.1 |
| PAA 5100 | 6.00 | 96.4 |
| PAA 100000 | 6.00 | 96.0 |
| PAA 100000 | 10.0 | 95.7 |
| PAA 450000 | 6.00 | 97.7 |
| PVP-co-PAA | 4.00 | 89.2 |
| PSS-co-PMA | 4.00 | 97.0 |
| PAAm-co-PAA 200K | 4.00 | 97.3 |
| PAAm-co-PAA 200K | 4.00 | 97.0 |
| PAAm-co-PAA 520K | 3.55 | 92.0 |
| PAAm-co-PAA 520K | 4.00 | 93.7 |
| polyDADMAC 28000 | 4.00 | 94.2 |
| polyDADMAC 165000 | 4.00 | 92.2 |
| polyDADMAC 941000 | 4.00 | 92.8 |

**Table S 1.** Adsorption rates from TOC analysis of investigated polymers on C-S-H synthesized *via* titration at pH 12.

## 9. Titration curves in presence of polymers at pH 12

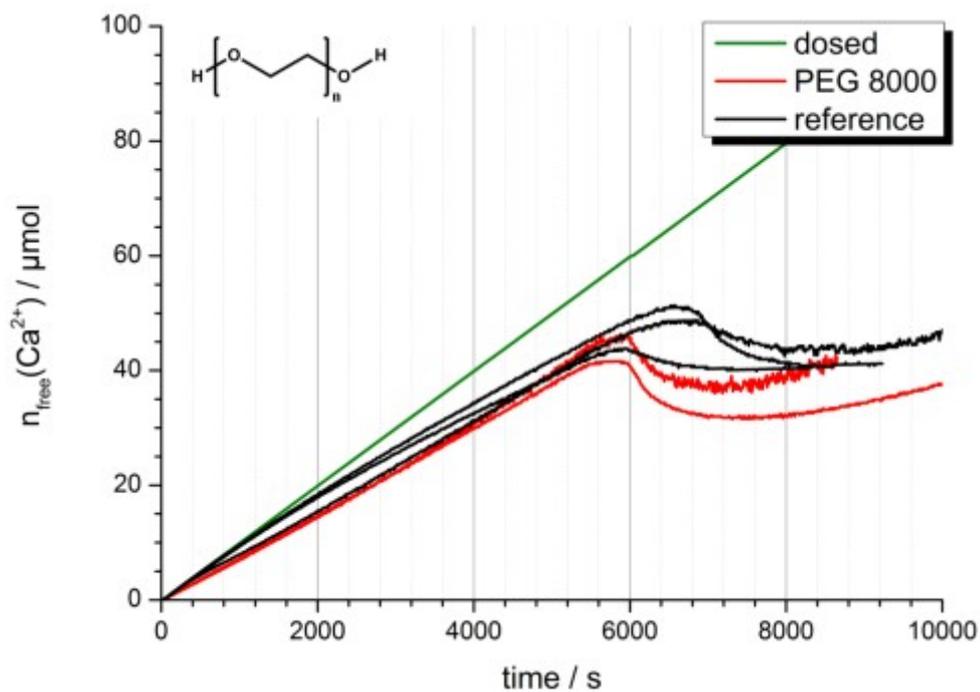

**Figure S 8.** PEG 8000, pH 12.

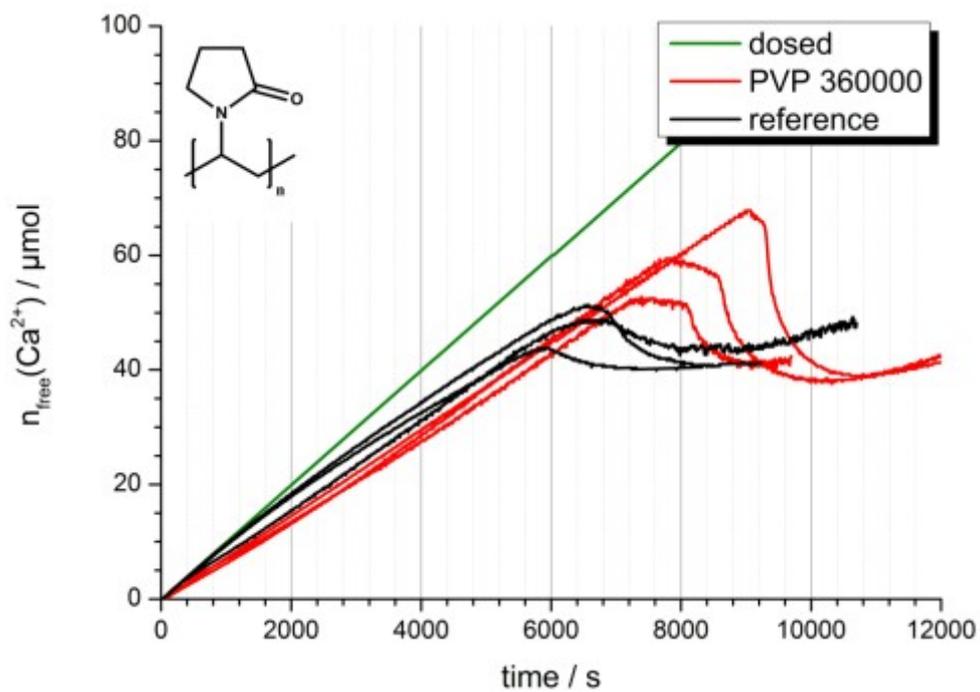

**Figure S 9.** PVP 360000, pH 12.

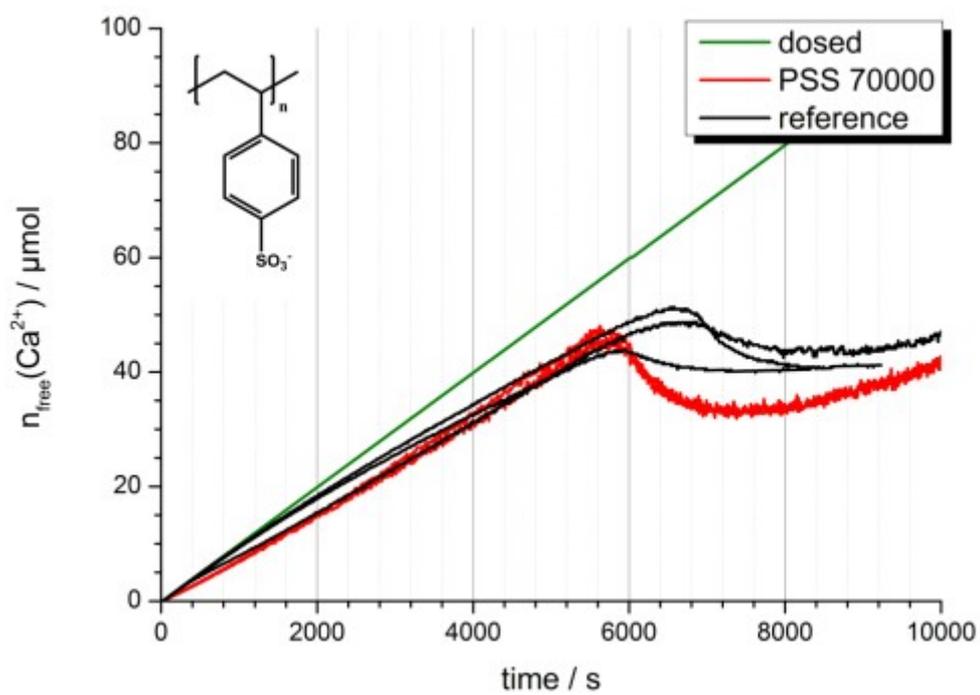

**Figure S 10.** PSS 70000, pH 12.

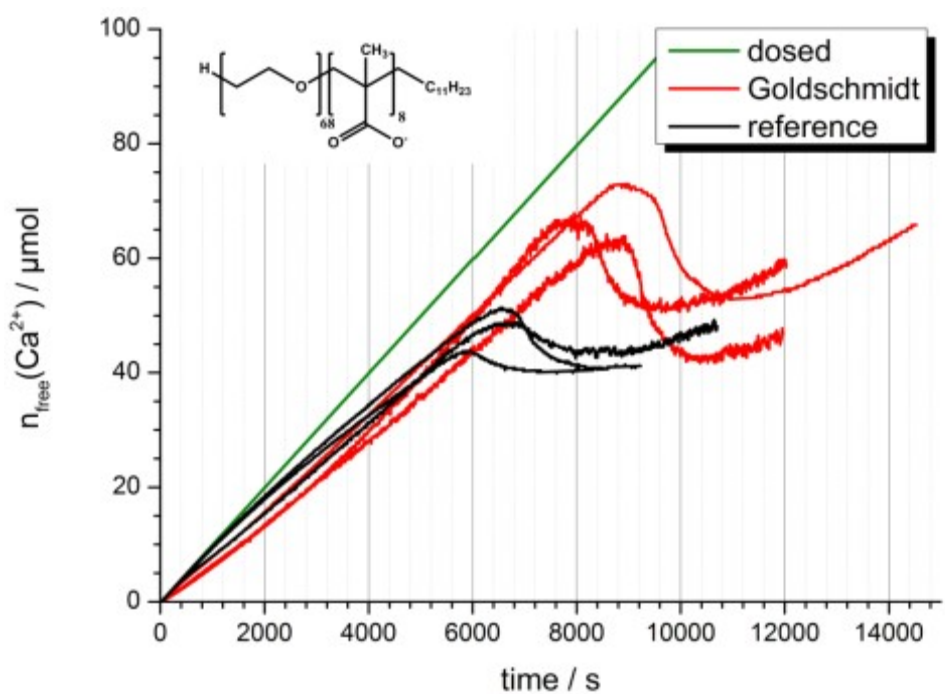

**Figure S 11.** Goldschmidt, pH 12.

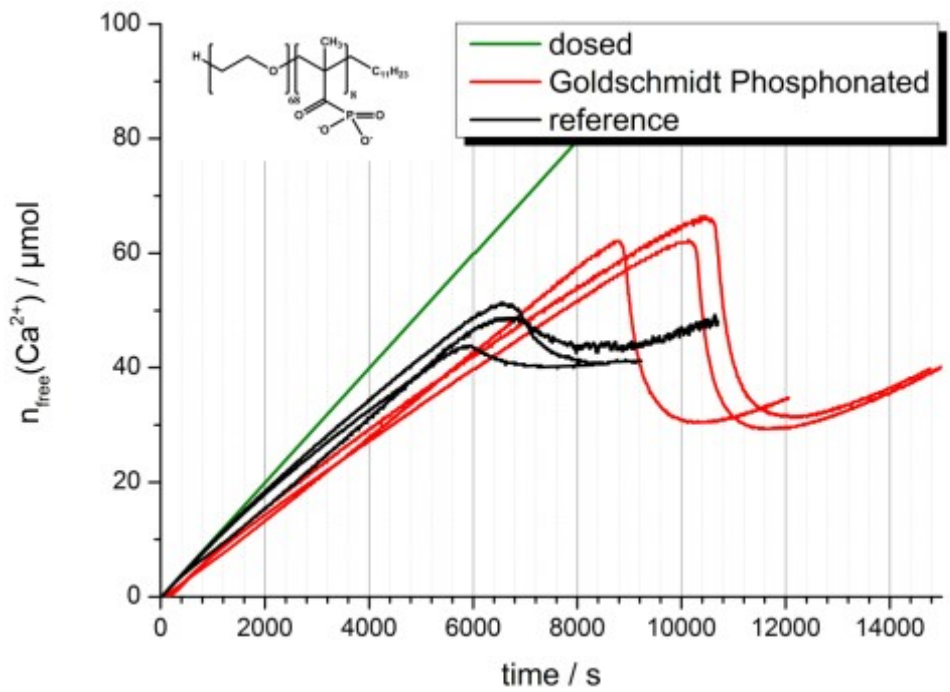

**Figure S 12.** Goldschmidt phosphonated, pH 12.

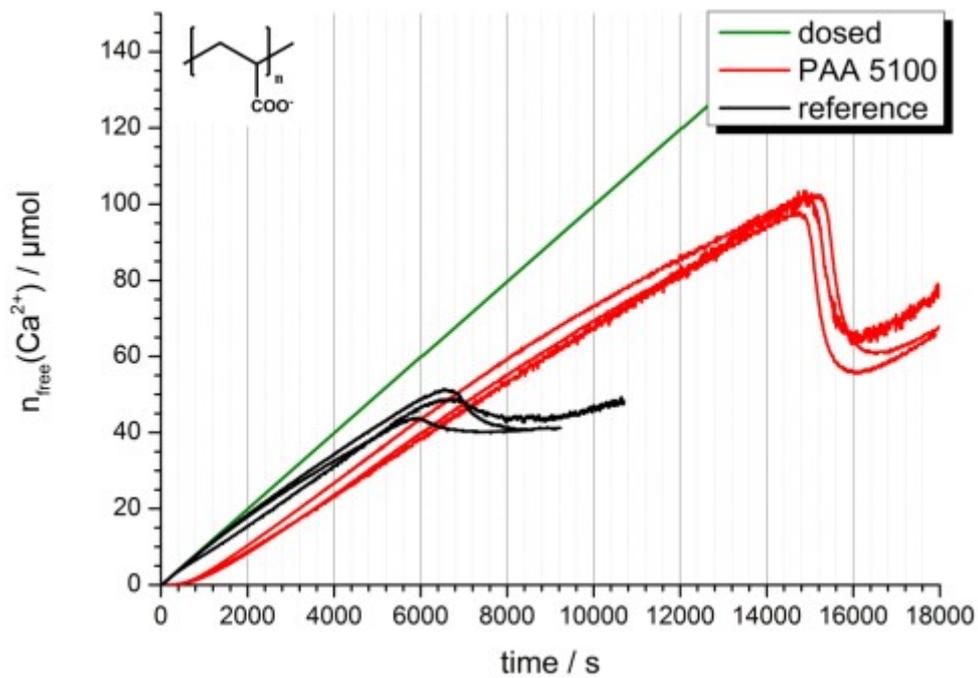

**Figure S 13.** PAA 5100, pH 12.

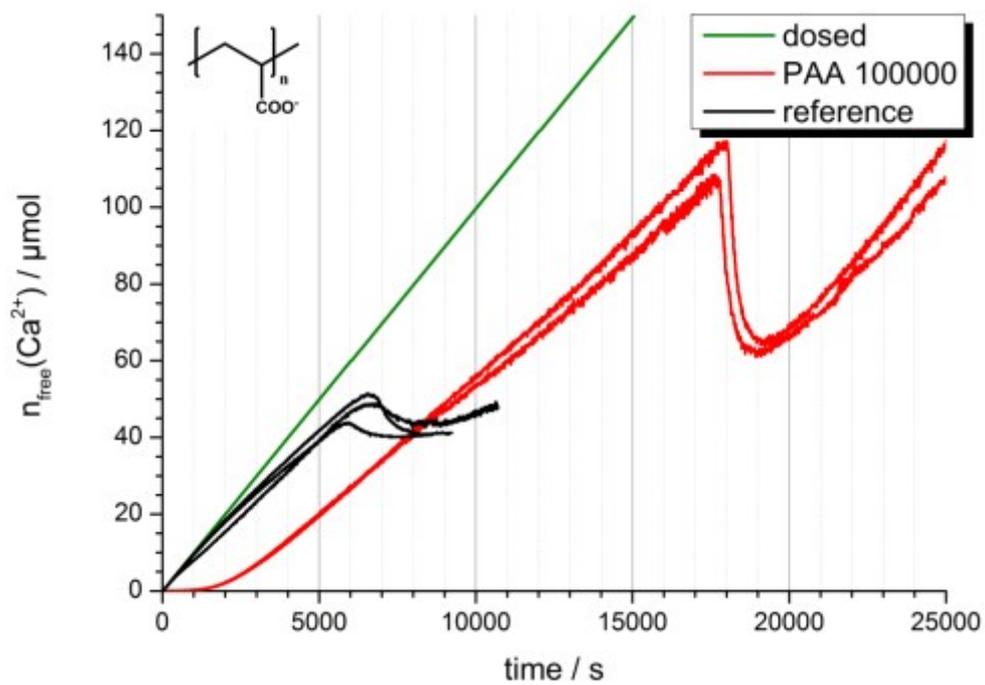

**Figure S 14.** PAA 100000, pH 12.

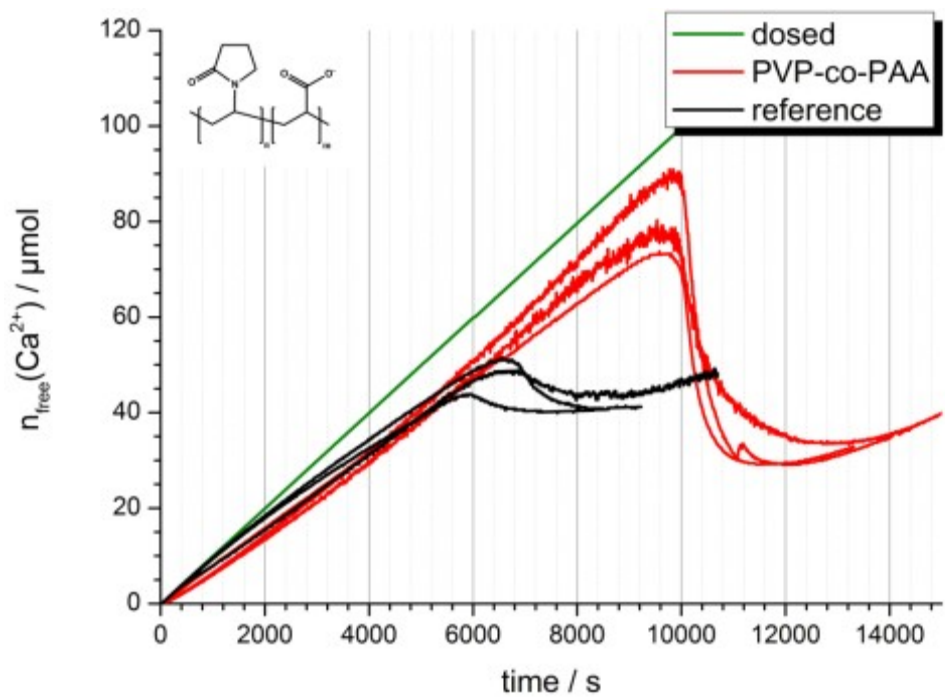

**Figure S 15.** PVP-co-PAA, pH 12.

**Figure S 16**. PSS-co-PMA, pH 12.

**Figure S 17.** PAAm-co-PAA 200000, pH 12.

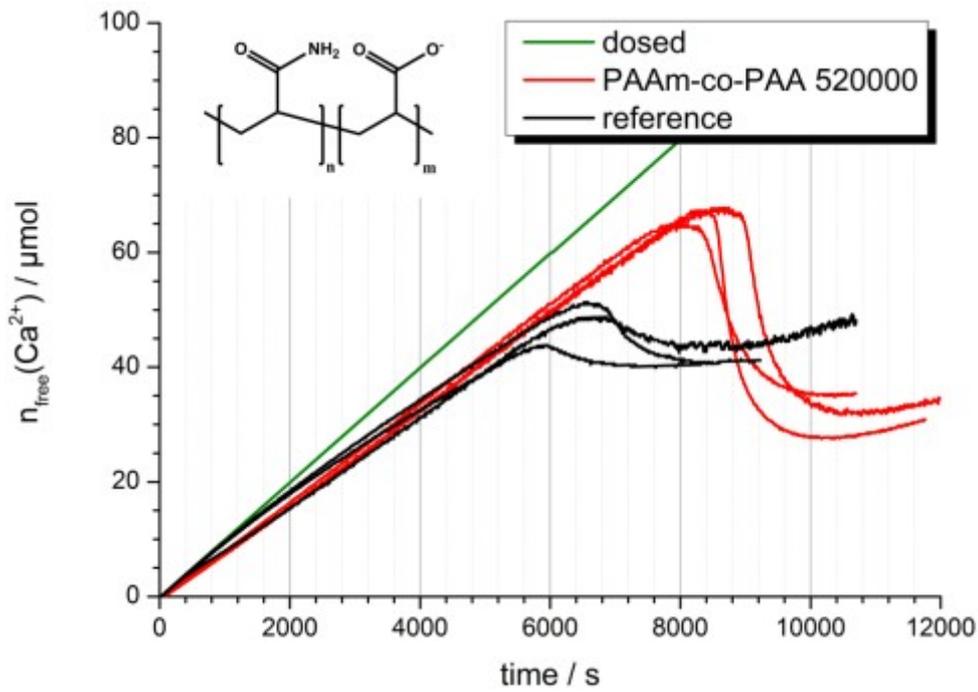

**Figure S 18.** PAAm-co-PAA 520000, pH 12.

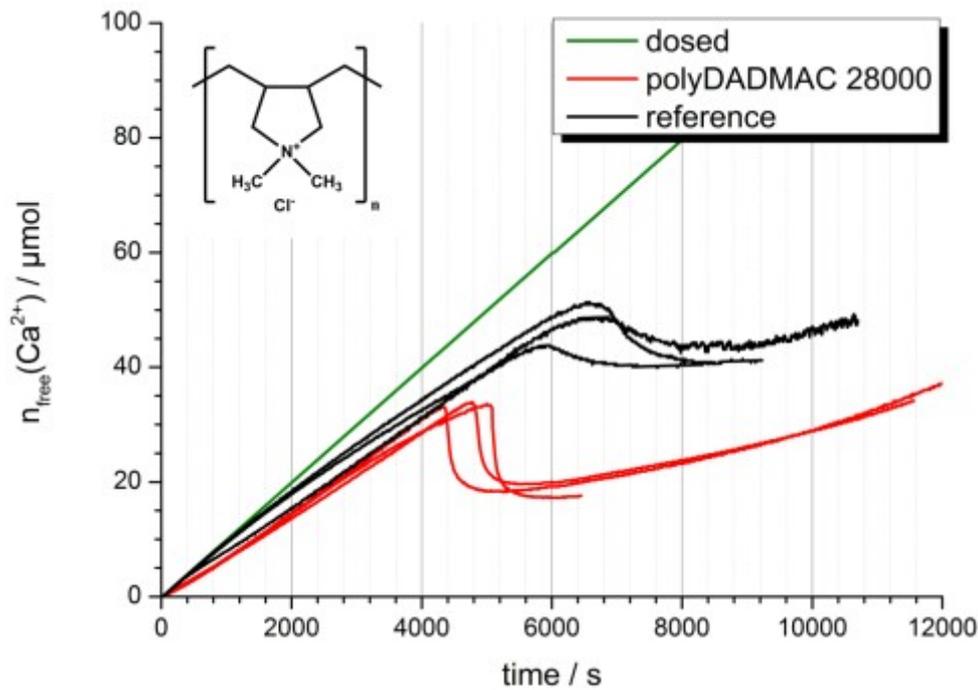

**Figure S 19.** polyDADMAC 28000, pH 12.

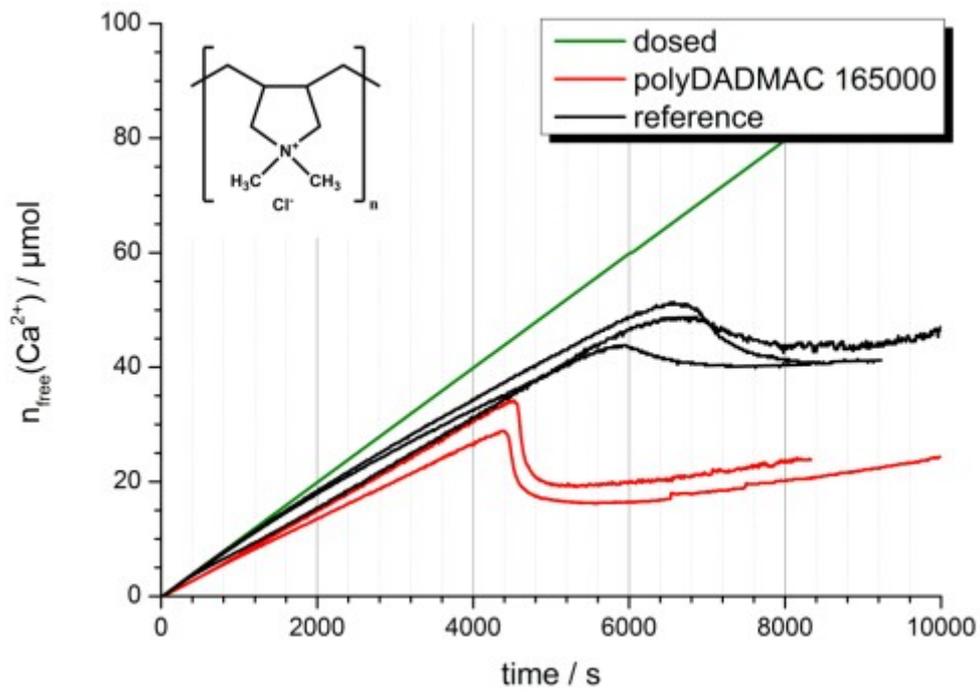

**Figure S 20**. polyDADMAC 165000, pH 12.

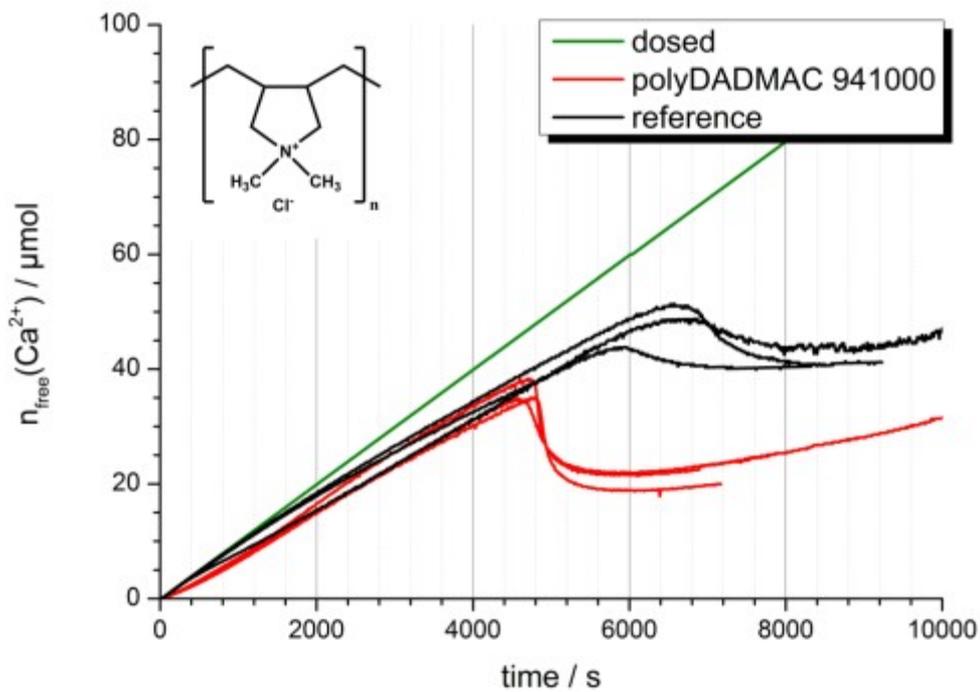

**Figure S 21**. polyDADMAC 941000, pH 12.

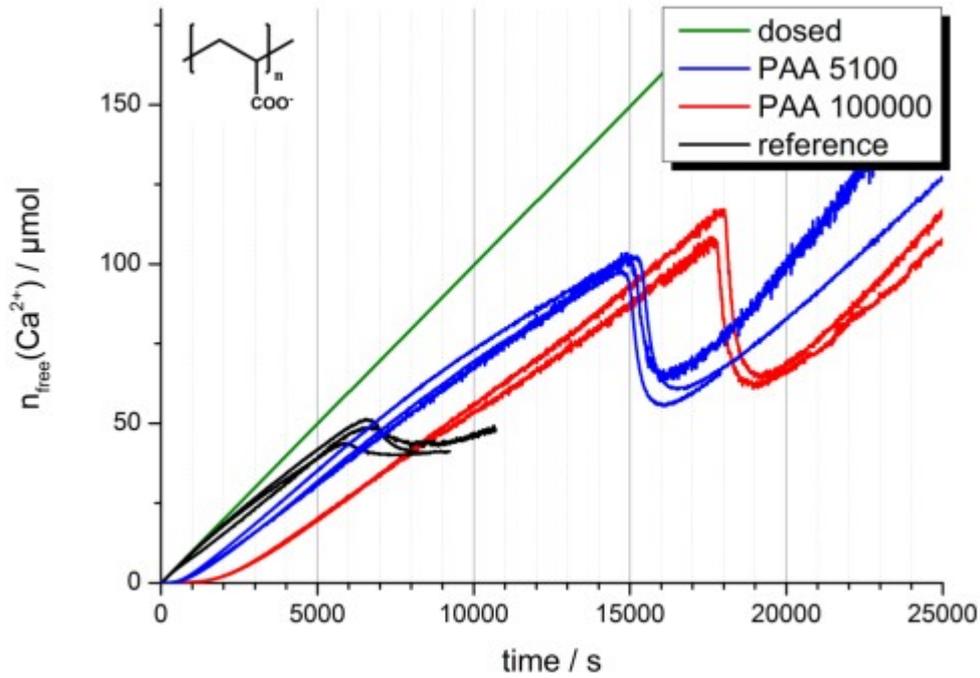

**Figure S 22.** Influence of different PAA chain lengths on the nucleation of C-S-H at pH 12.

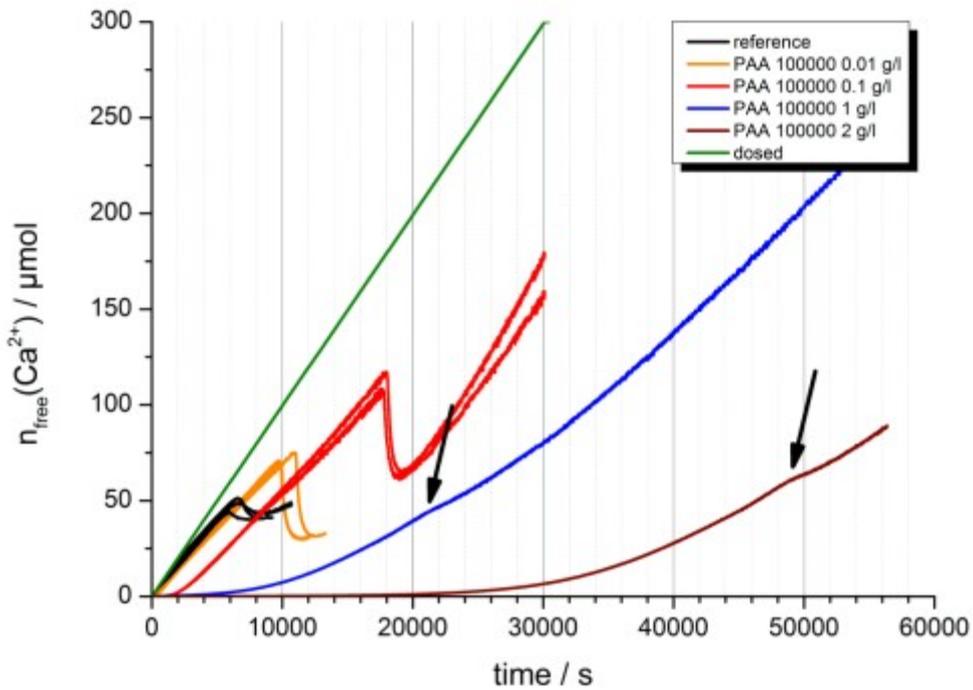

**Figure S 23.** Influence of different concentrations of PAA 100000 on the nucleation of C-S-H at pH 12. The arrows indicate the weakly visualized nucleation points for high PAA concentrations.

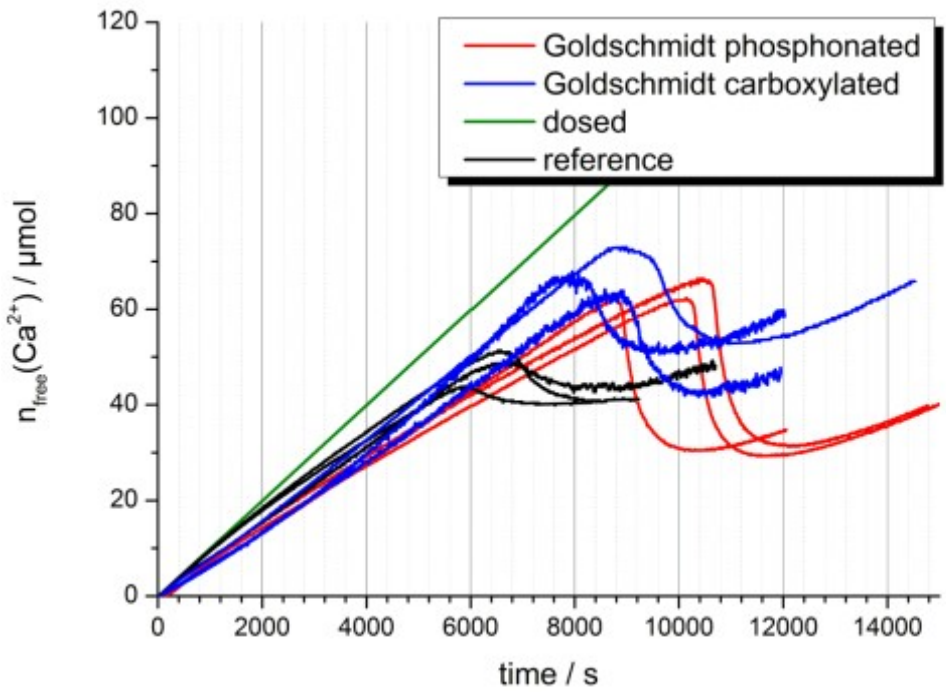

**Figure S 24.** Comparison of the efficiency of different functional groups on the nucleation of C-S-H at pH 12.

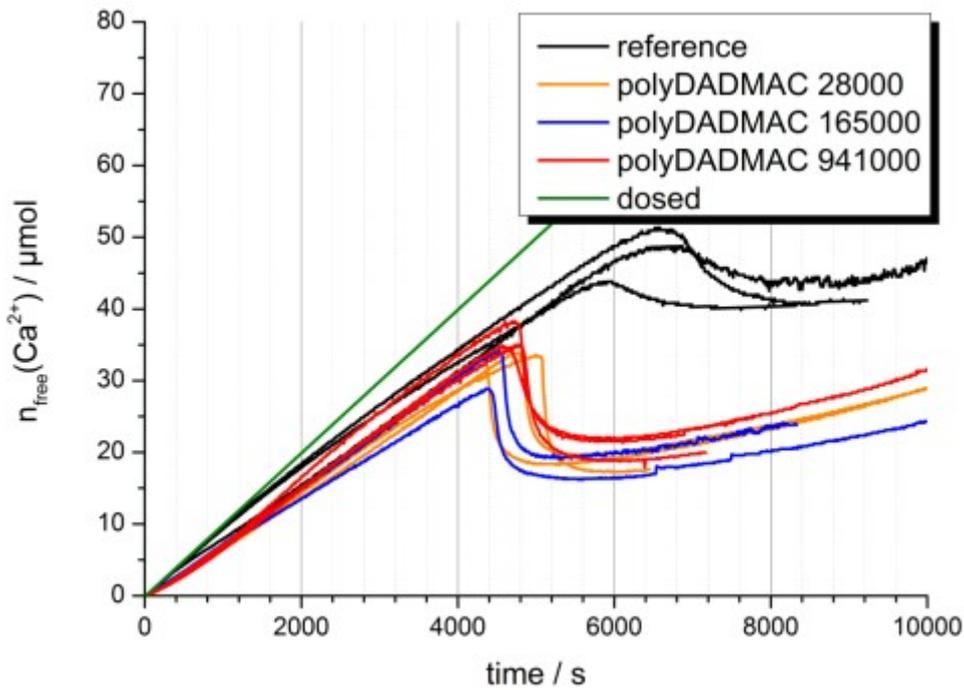

**Figure S 25.** Influence of different polyDADMAC chain lengths on the nucleation of C-S-H at pH 12.

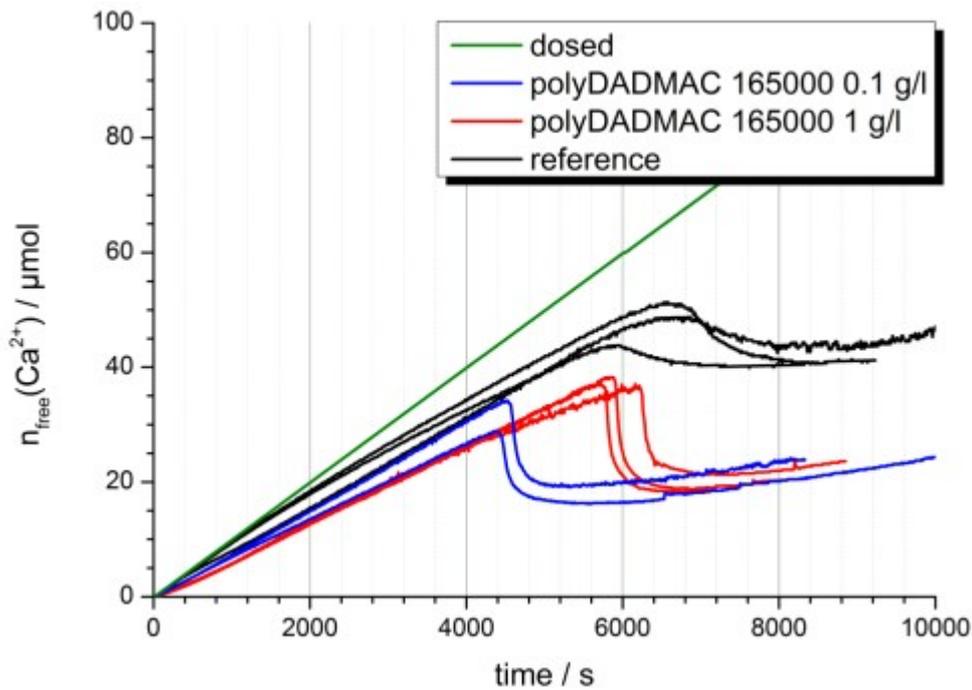

**Figure S 26.** Influence of different concentrations of polyDADMAC 165000 on the nucleation of C-S-H at pH 12.

## 10. Adsorption of polymers on C-S-H at pH 13

| Polymer | $V_{added}$(CaCl$_2$) [ml] | Adsorbed on C-S-H [%] |
|---|---|---|
| PEG 8000 | 2.20 | 87.3 (92.5) |
| PVP 360000 | 2.20 | 84.0 (82.2) |
| PVA 9000 | 2.20 | 89.2 (83.0) |
| PSS 70000 | 2.20 | 92.2 (94.9) |
| GS-P | 2.20 | 87.2 (86.1) |
| GS | 2.20 | 86.2 (87.2) |
| PAA 5100 | 2.20 | 93.9 (97.1) |
| PAA 5100 | 3.50 | 94.5 |
| PAA 100000 | 3.50 | 86.5 |
| PAA 450000 | 3.50 | 95.6 |
| PVP-co-PAA | 2.20 | 90.1 (89.2) |
| PSS-co-PMA | 2.20 | 94.3 (97.0) |
| PAAm-co-PAA 200K | 2.20 | 96.7 (97.3) |
| PAAm-co-PAA 200K | 3.50 | 96.1 |
| PAAm-co-PAA 520K | 2.20 | 91.1 (93.7) |
| polyDADMAC 28000 | 2.20 | 83.7 (94.2) |
| polyDADMAC 165000 | 2.20 | 93.5 (92.2) |
| polyDADMAC 941000 | 2.20 | 80.6 (92.8) |

**Table S 2.** Adsorption rates from TOC analysis of investigated polymers on C-S-H synthesized via titration at pH 13. Values in brackets are TOC results from pH 12 with equal supersaturation with respect to C-S-H.

## 11. Titration curves in presence of polymers at pH 13

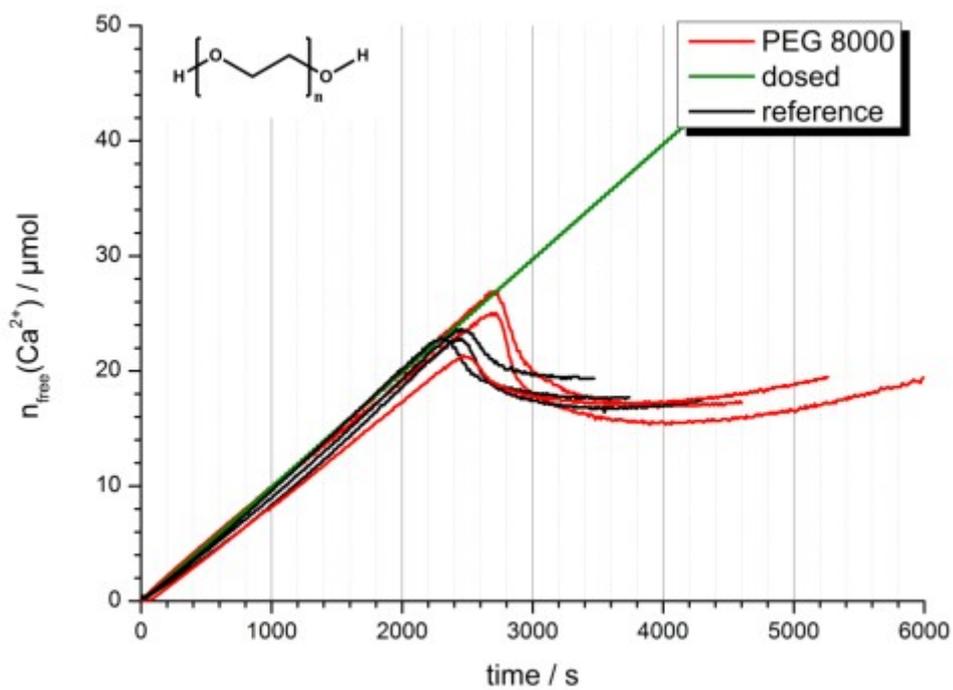

**Figure S 27.** PEG 8000, pH 13.

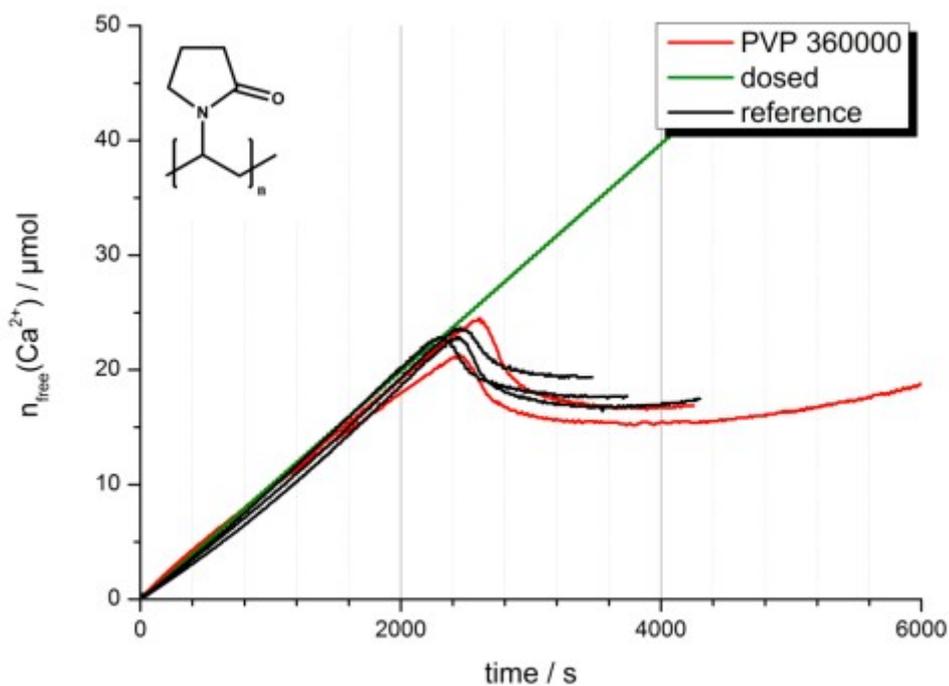

**Figure S 28**. PVP 360000, pH 13.

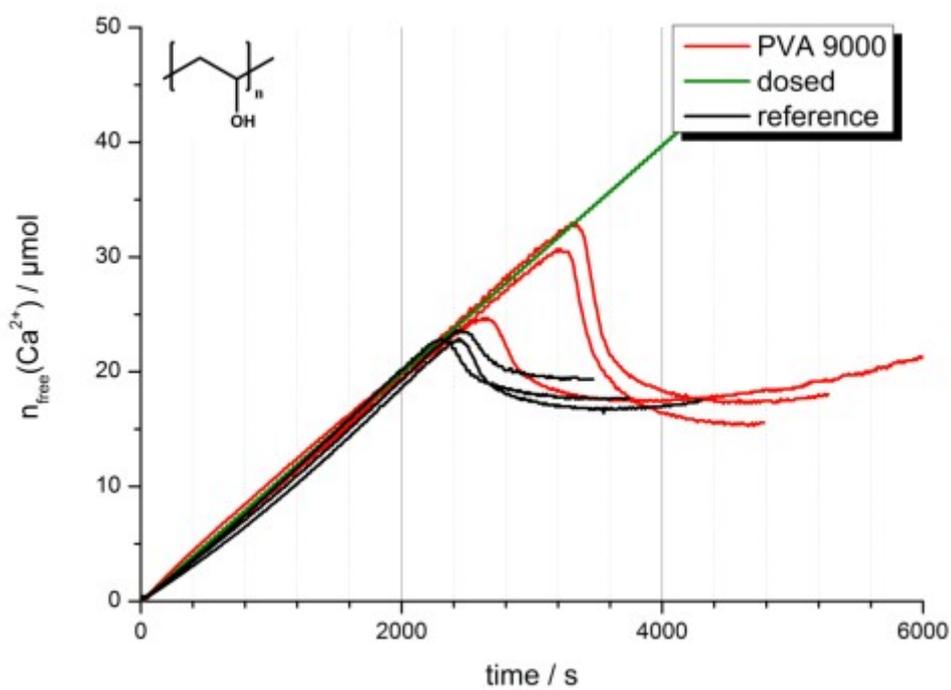

**Figure S 29.** PVA 9000 - 10000, pH 13.

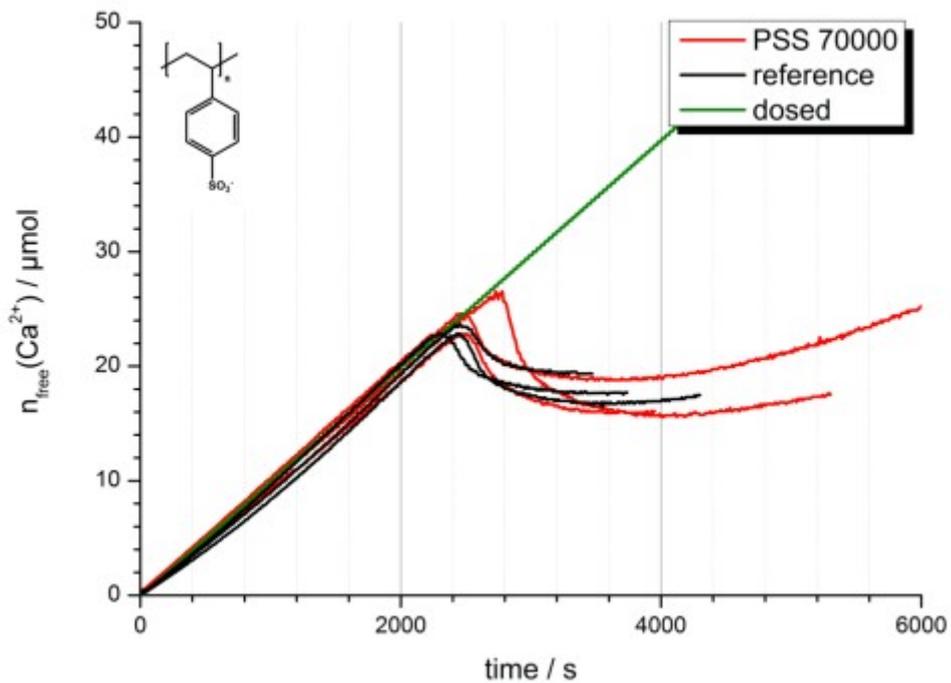

**Figure S 30.** PSS 70000, pH 13.

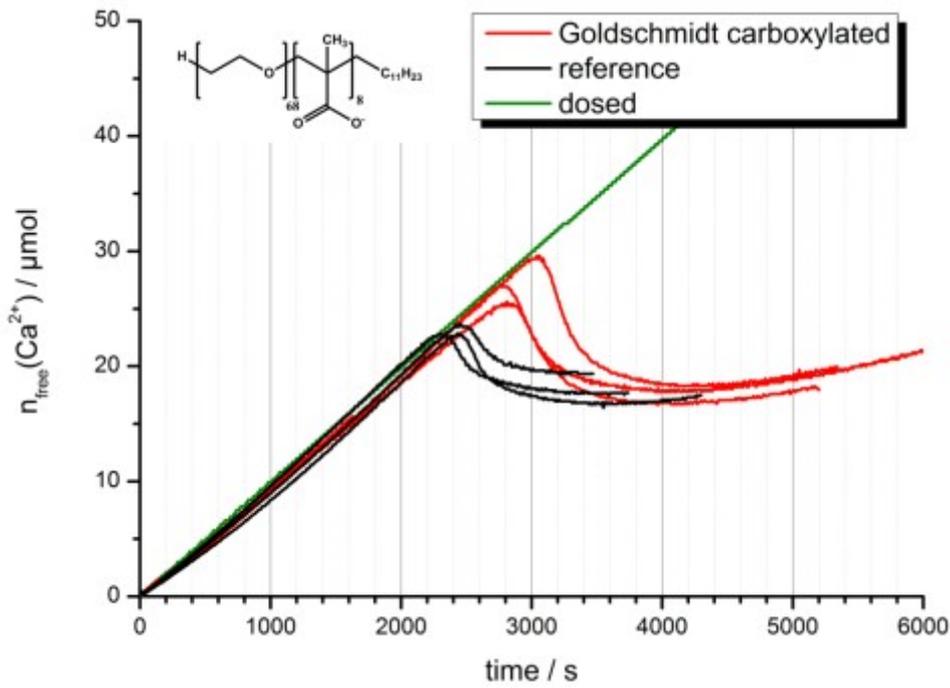

**Figure S 31**. Goldschmidt, pH 13.

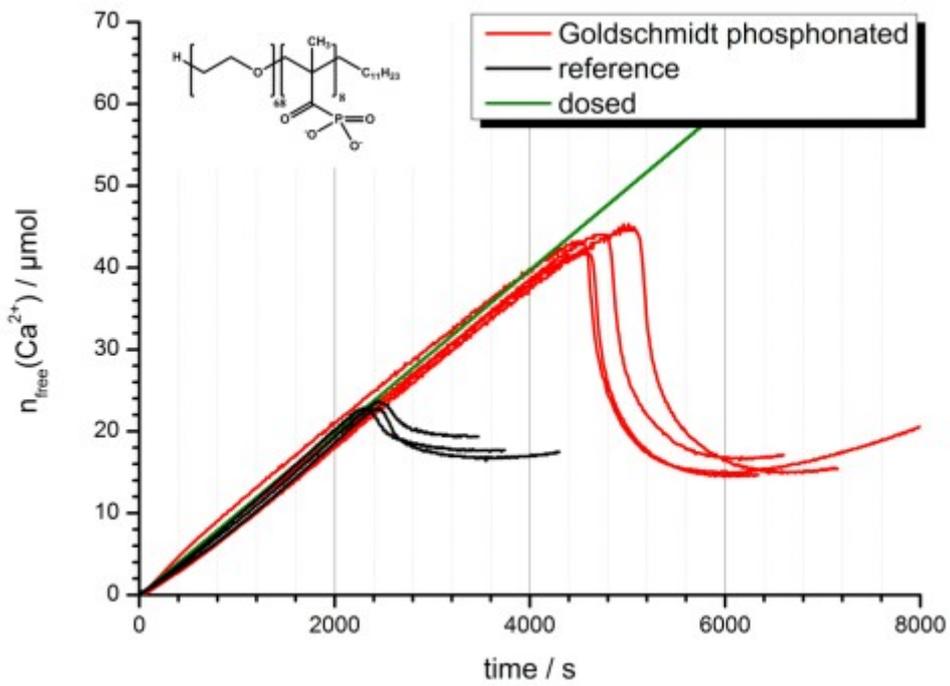

**Figure S 32.** Goldschmidt phosphonated, pH 13.

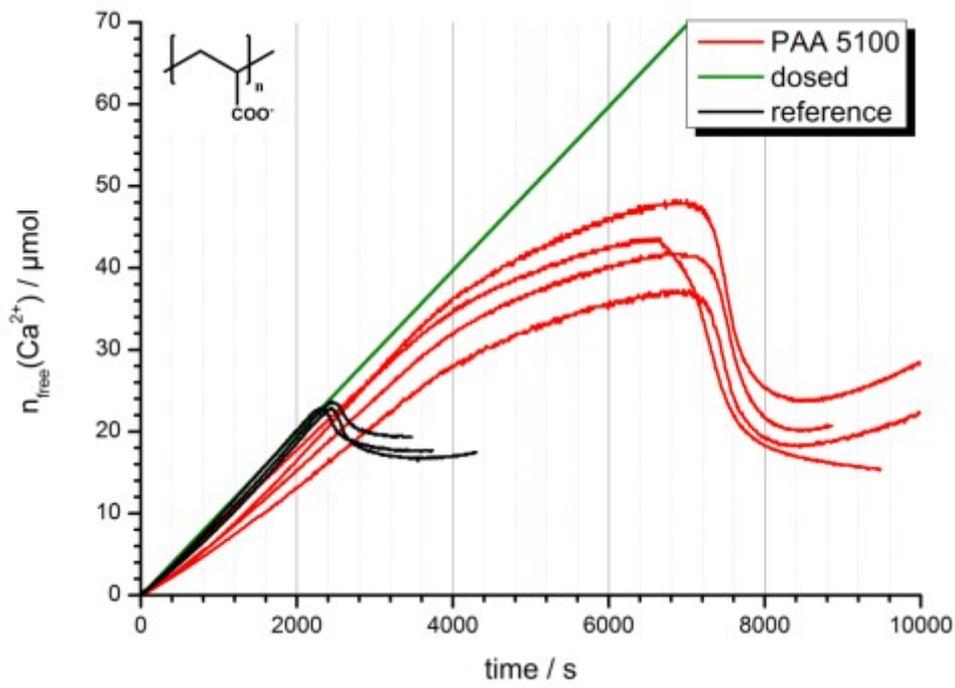

**Figure S 33.** PAA 5100, pH 13.

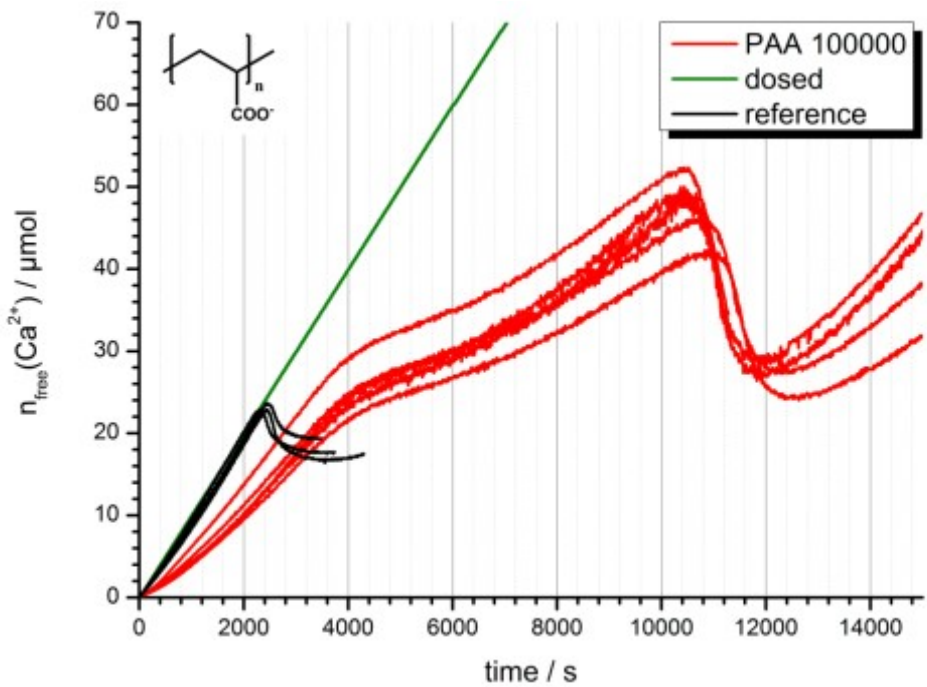

**Figure S 34.** PAA 100000, pH 13.

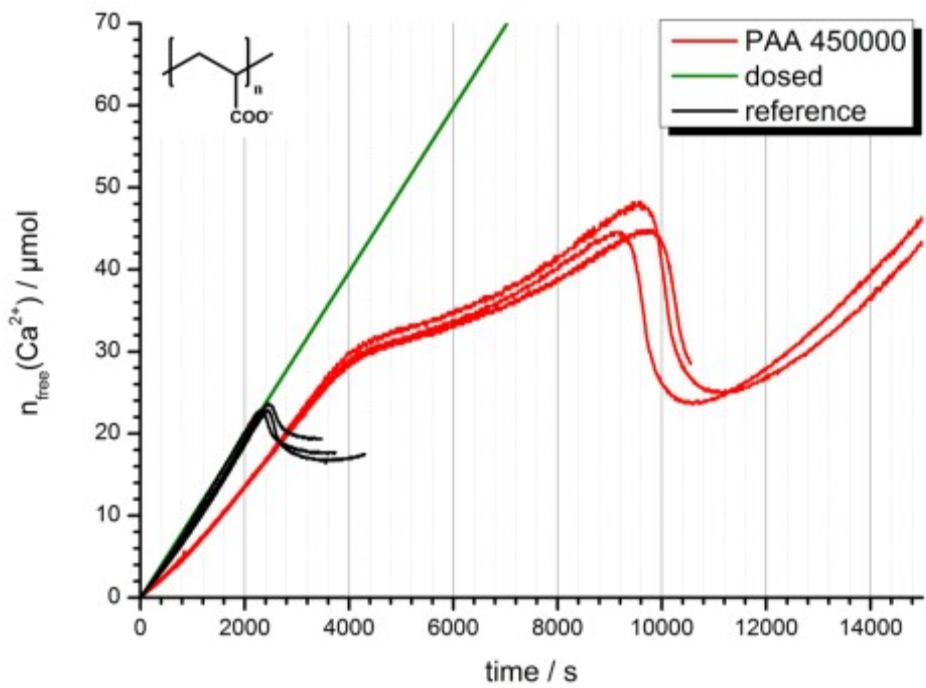

**Figure S 35**. PAA 450000, pH 13.

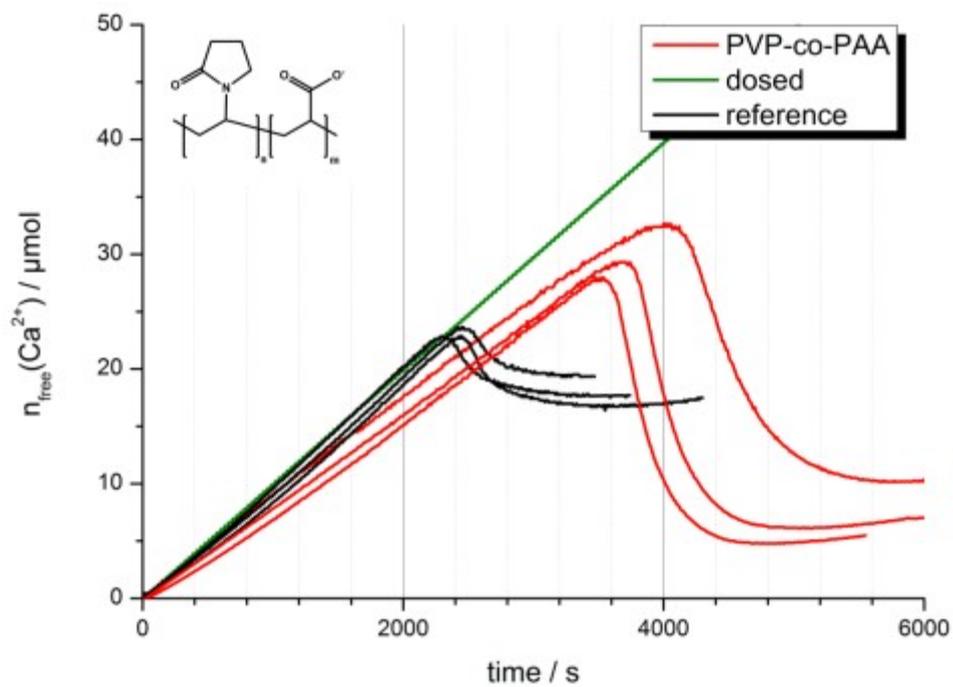

**Figure S 36.** PVP-co-PAA, pH 13.

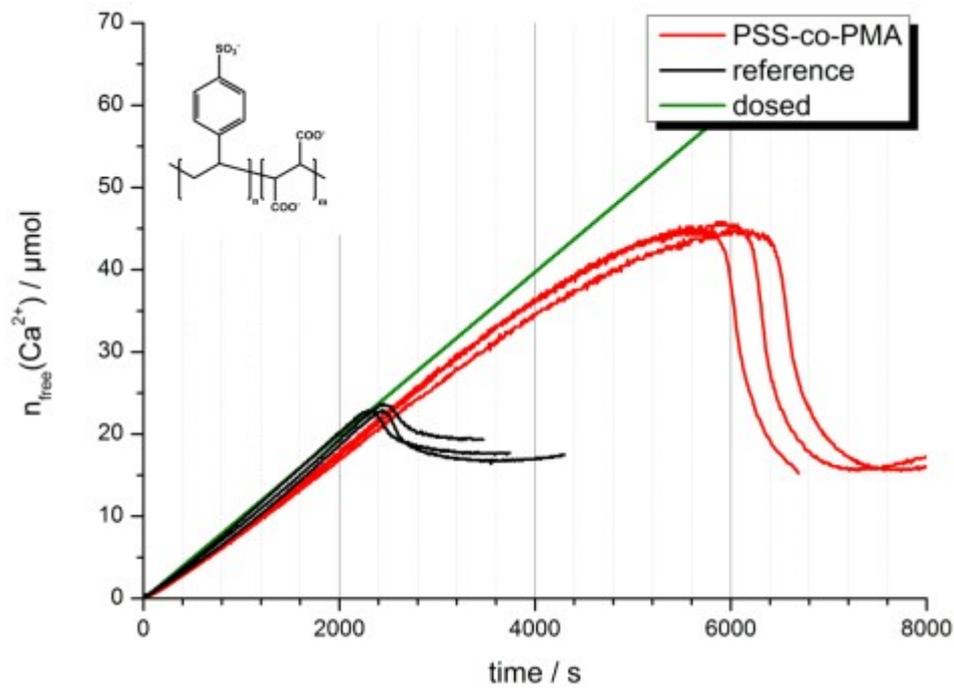

**Figure S 37.** PSS-co-PMA, pH 13.

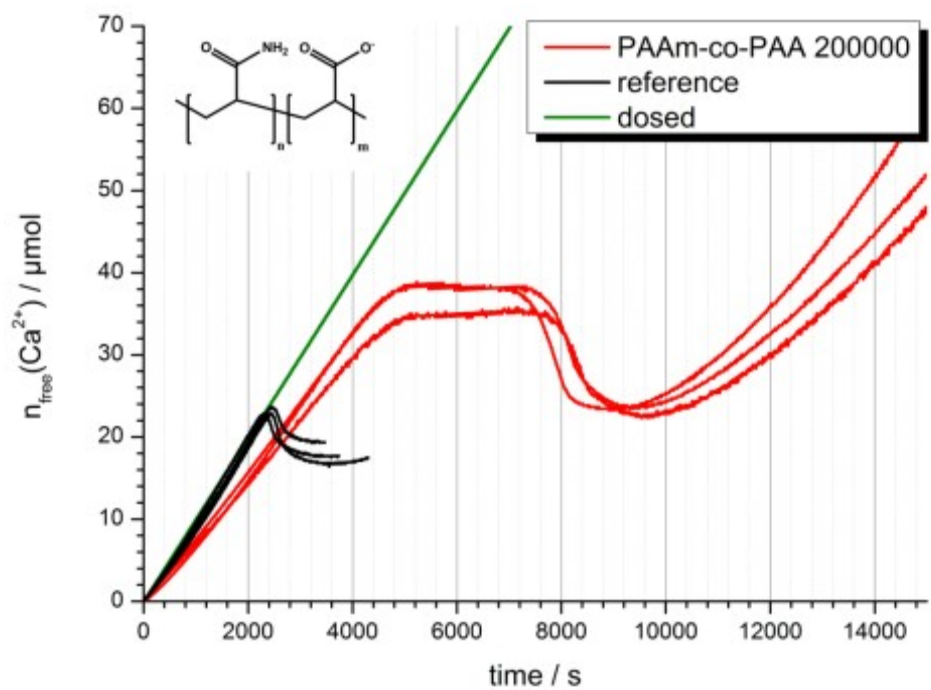

**Figure S 38.** PAAm-co-PAA 200000, pH 13.

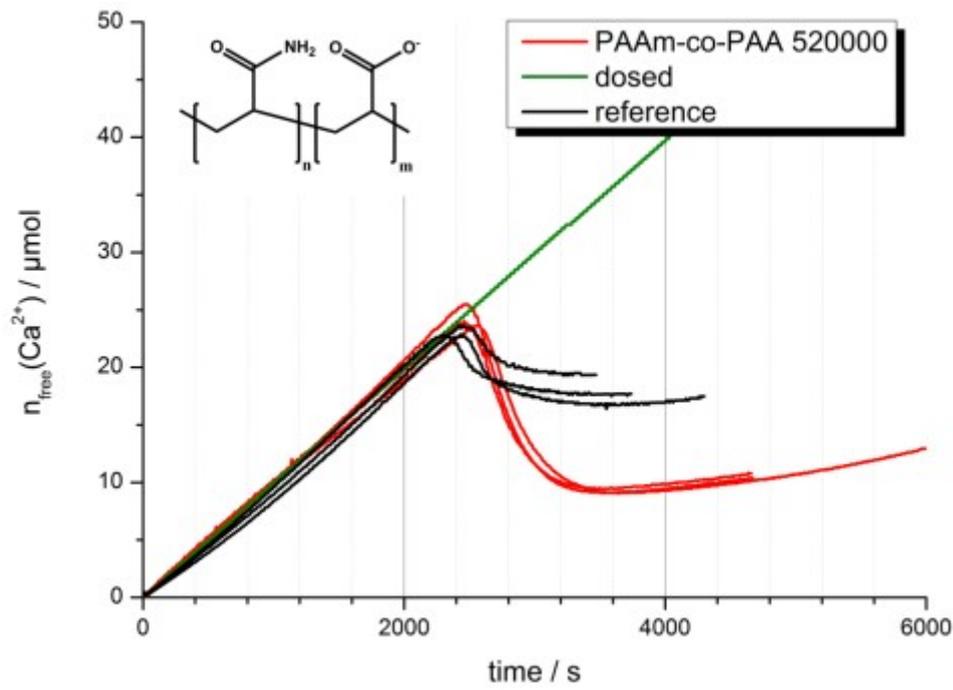

**Figure S 39.** PAAm-co-PAA 520000, pH 13.

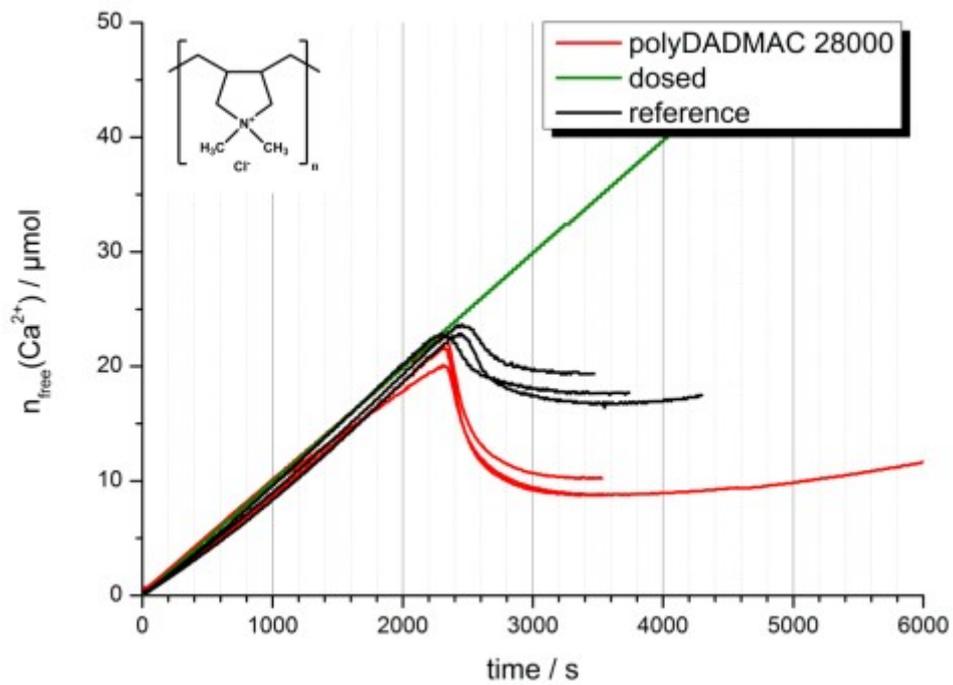

**Figure S 40.** polyDADMAC 28000, pH 13.

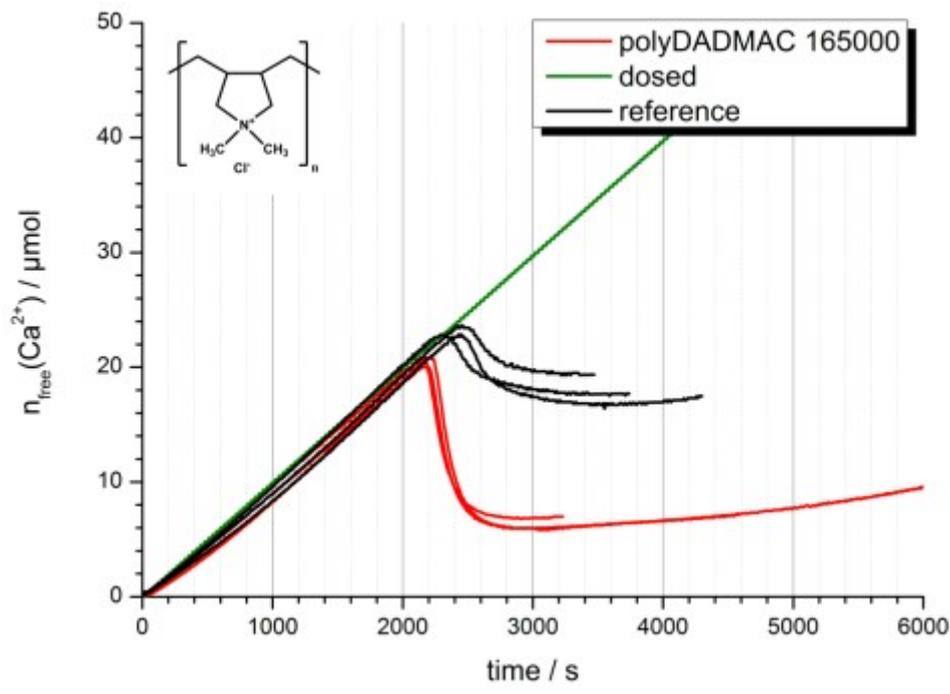

**Figure S 41**. polyDADMAC 165000, pH 13.

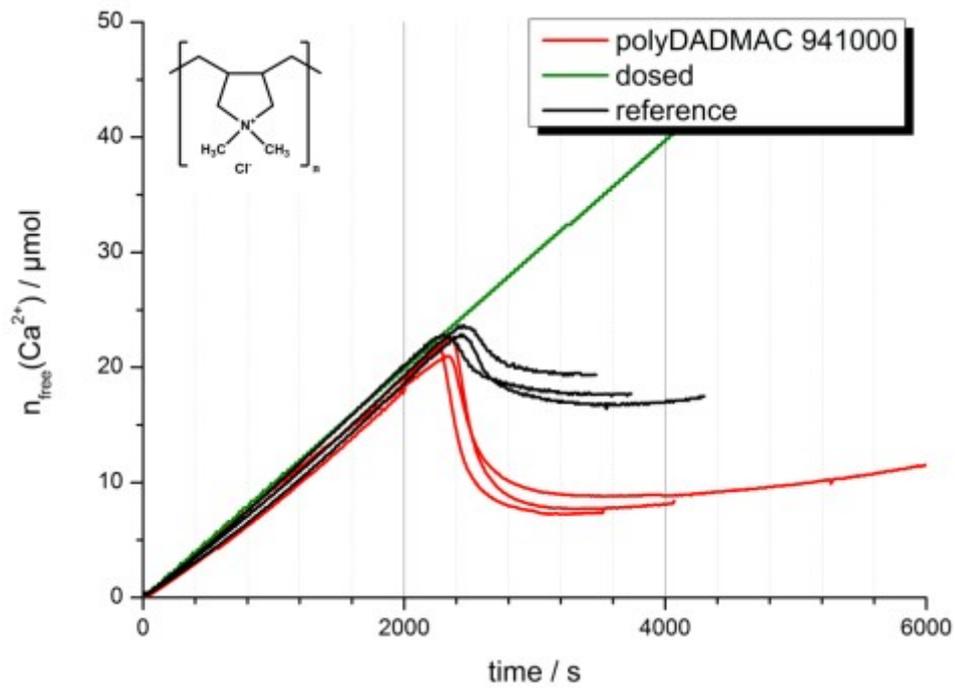

**Figure S 42.** polyDADMAC 941000, pH 13.

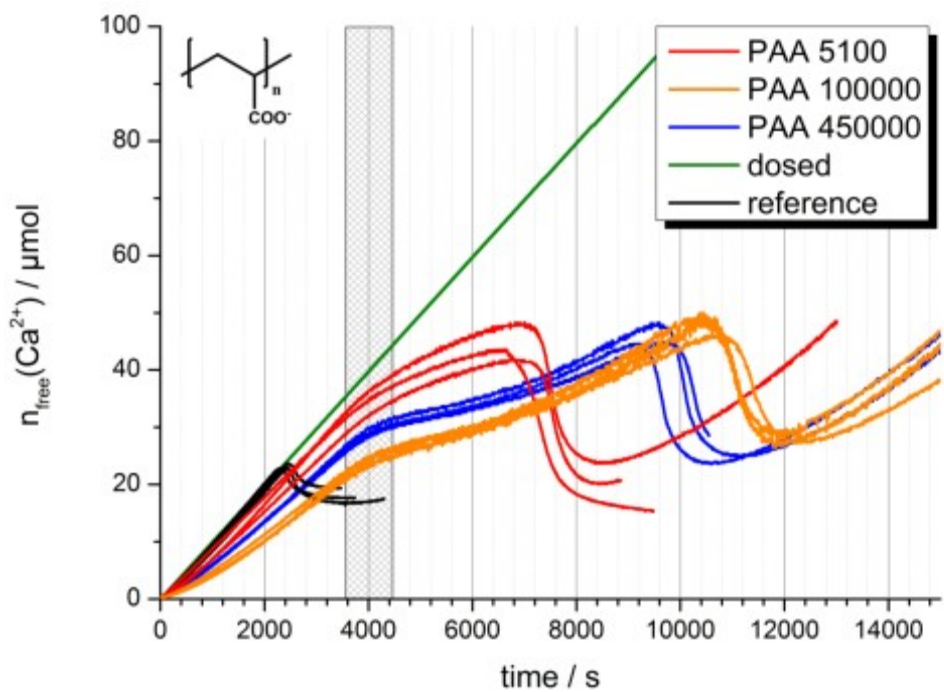

**Figure S 43**. Influence of different PAA chain lengths on the nucleation of C-S-H at pH 13. The shaded area indicates the region where bending occurs.

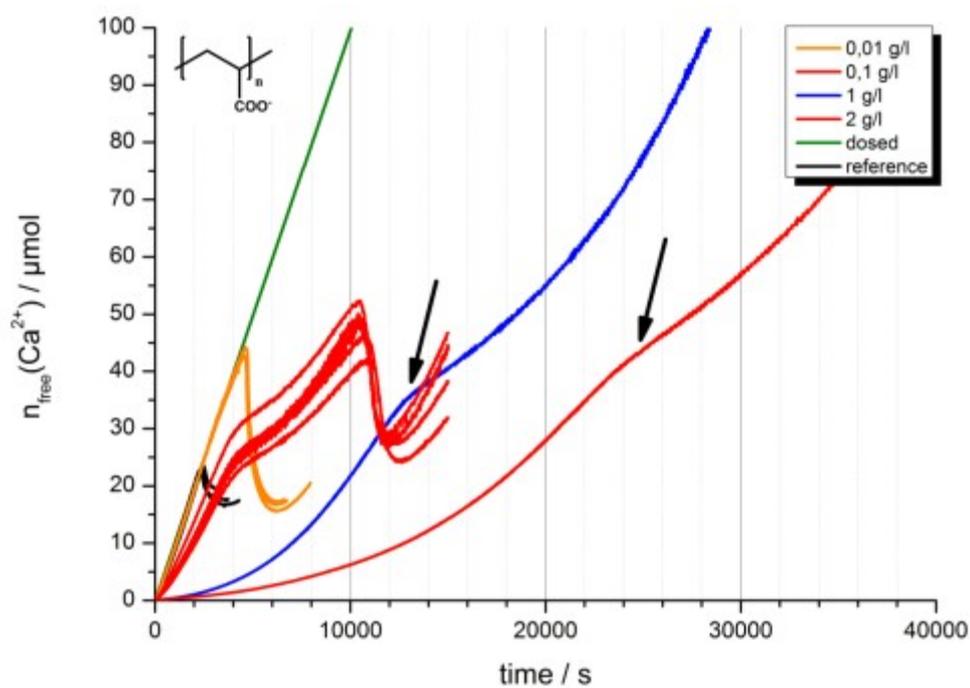

**Figure S 44.** Influence of different PAA 100000 concentrations on the nucleation of C-S-h at pH 13.

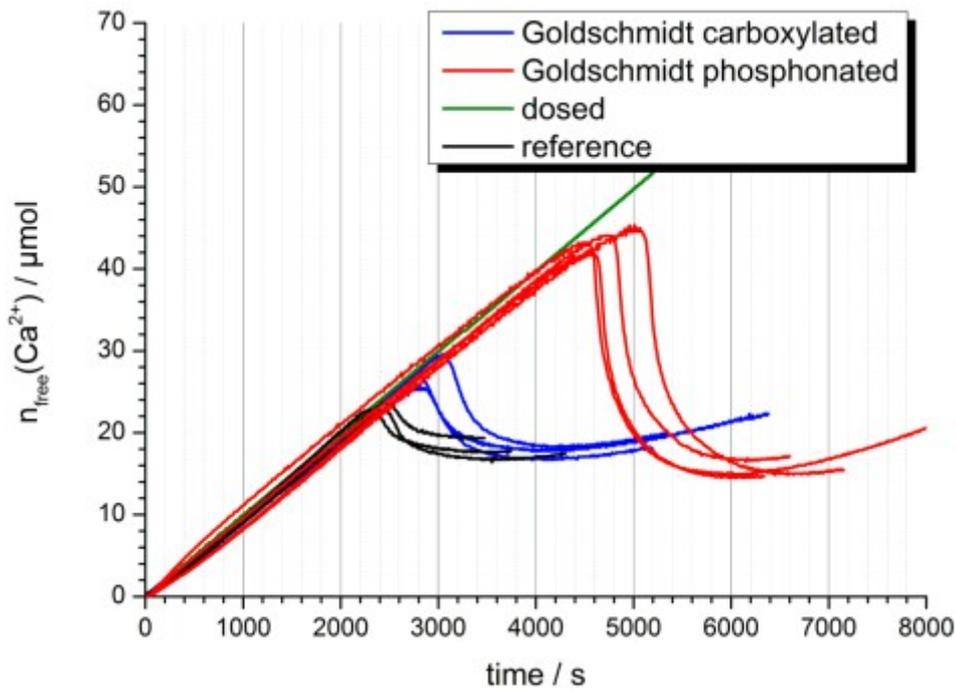

**Figure S 45.** Comparison of the efficiency of different functional groups on the nucleation of C-S-H at pH 13.

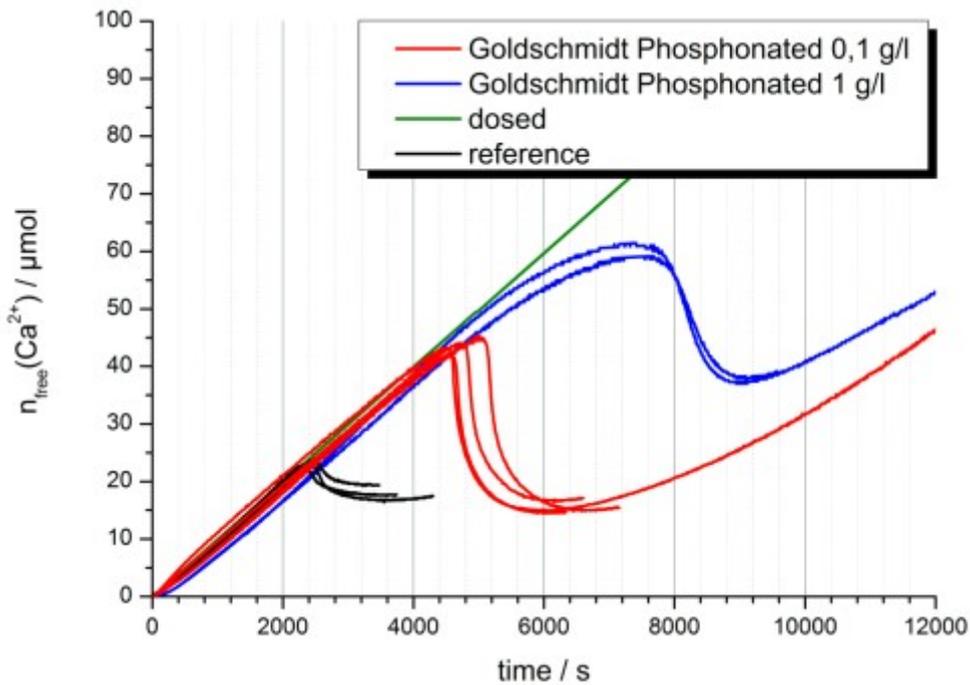

**Figure S 46**. Influence of different concentrations of Goldschmidt phosphonated on the nucleation of C-S-H at pH 13.

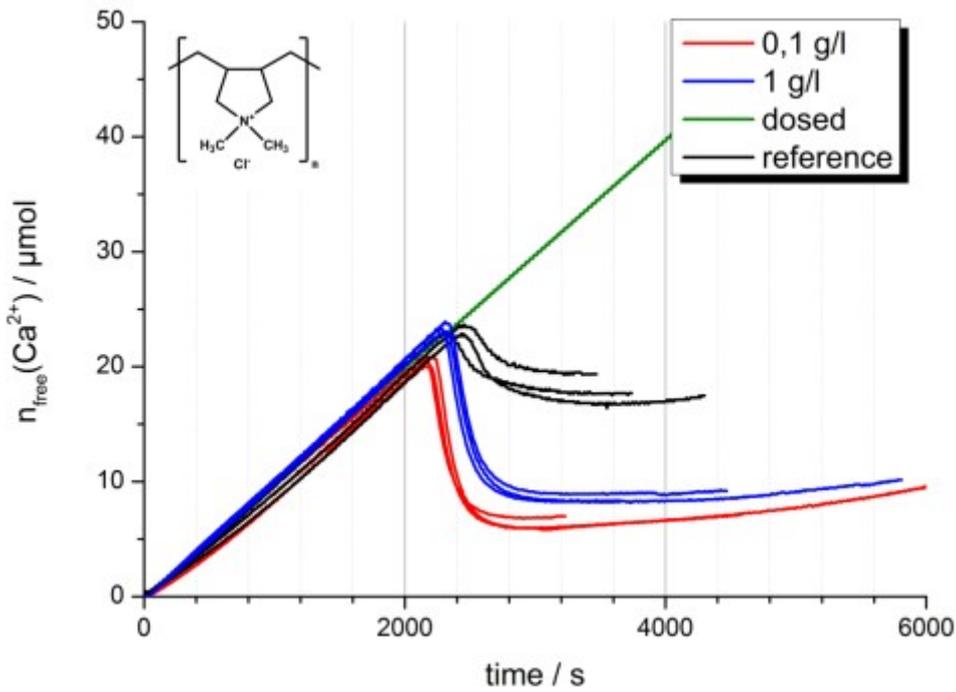

**Figure S 47**. Influence of different concentrations of polyDADMAC 165000 on the nucleation of C-S-H at pH 13.

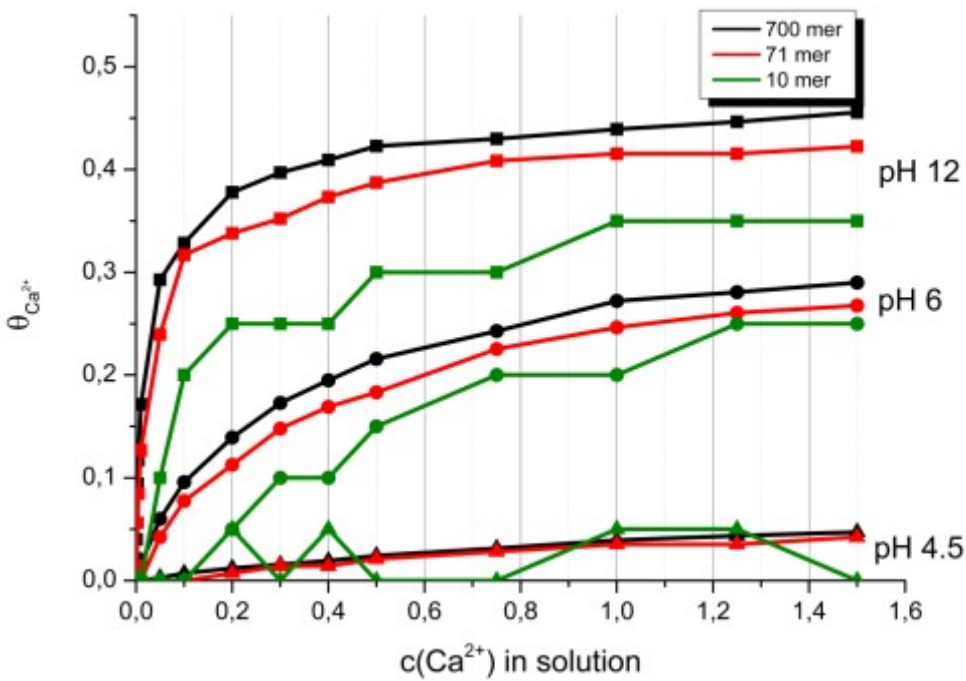

**Figure S 48**. Monte Carlo simulations to confirm the trend of higher $Ca^{2+}$ adsorption capacity with increasing chain length of acrylic acid. Poly (acrylic acids) with chain lengths of 10, 71 and 700 monomers (71 corresponding to PAA 5100) were added in a simulation box together with a varying amount of calcium ions (adjusted via the chemical potential of the 2:1 salt $CaCl_2$). This value was then compared to the number of calcium ions in the very same simulation box in absence of polymers, while the difference gave the amount of adsorbed calcium on the polymers..

## 12. Quantitative Determination of the Ca$^{2+}$-binding capacity of polymers

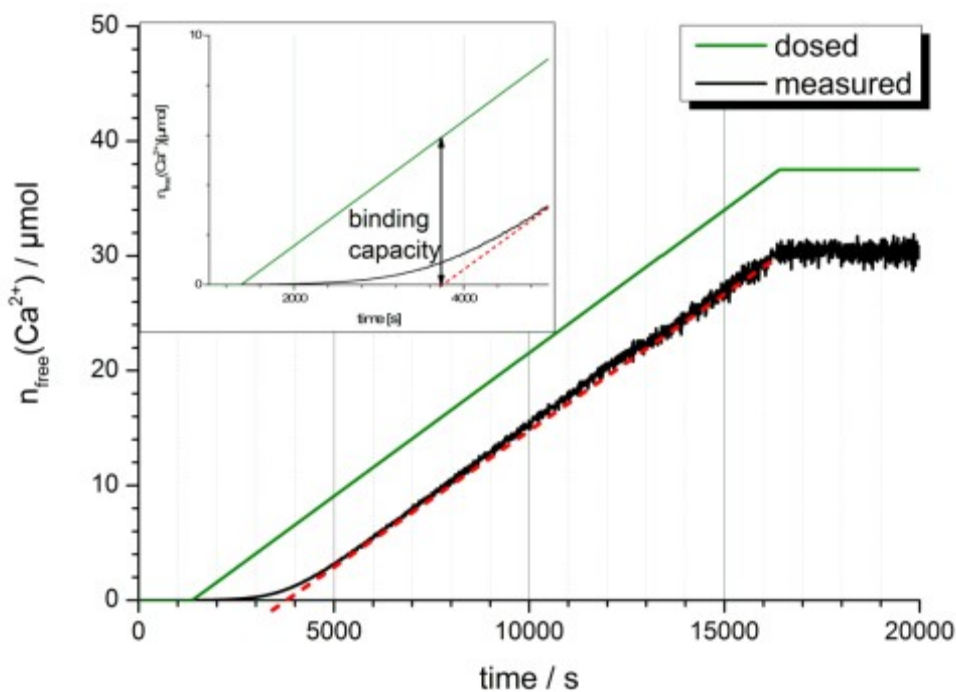

**Figure S 49.** Quantitative determination of the Ca$^{2+}$ binding of PAA 5100. CaCl$_2$ is dosed into a polymer solution at pH 12 and the calcium potential is measured by an ion selective electrode. The difference between the dosed and measured amount of free calcium is bound by the polymer.

| Polymer | Functional side group | M$_W$ [g/mol] | F$_R$ at pH 12 | F$_R$ at pH 13 | ß at pH 12 | ß at pH 13 | Missing Ca$^{2+}$ at nucleation point, pH 12 | Missing Ca$^{2+}$ at nucleation point, pH 13 | Bound Ca$^{2+}$ at pH 12 per interesting monomer |
|---|---|---|---|---|---|---|---|---|---|
| reference | --- | --- | 1.00 | 1.00 | 24.5 | 19.5 | 15.1 | 0.5 | --- |
| PSS | S | 70000 | 0.90 | 1.07 | 22.4 | 21.4 | 10.9 | 0.9 | 0.13 |
| PVP | --- | 360000 | 1.32 | 1.04 | 28.3 | 19.8 | 20.2 | 1.4 | --- |
| PEG | --- | 8000 | 0.93 | 1.10 | 23.3 | 20.3 | 14.4 | 1.6 | --- |
| PVA | --- | 9000 | --- | 1.28 | --- | 26.1 | | 0.9 | --- |
| GS-P | P, C | 4000 | 1.52 | 1.96 | 29.3 | 38.3 | 34.1 | 3.2 | 0.07 |
| GS | C | 3950 | 1.32 | 1.21 | 33.4 | 23.7 | 16.5 | 1.4 | 0.03 |
| PAA | C | 5100 | 2.32 | 1.65 / 2.87 | 45.8 | 28.2 / 35.5 | 47.8 | 7.3 | 0.18 |
| PAA | C | 100000 | 2.75 | 1.66 / 4.42 | 50.8 | 19.6 / 31.1 | 64.2 | 15.3 | 0.53 |
| PAA | C | 450000 | --- | 1.66 / 3.97 | --- | 25.1 / 36.4 | | 10.9 | 0.34 |
| PVP-co-PAA | C | 96000 | 1.51 | 1.58 | 39.4 | 26.2 | 16.7 | 7.3 | 0.18 |
| PSS-co-PMA | S, C | 20000 | 1.93 | 2.50 | 49.7 | 38.4 | 15.9 | 13.7 | 0.51 |
| PAAm-co-PAA | C | 200000 | 2.09 | 1.99 / 3.06 | 38.0 | 31.7 | 49.5 | 11.5 | 0.27 |
| PAAm-co-PAA | C | 520000 | 1.30 | 1.04 | 32.3 | 20.3 | 16.3 | 0.6 | 0.05 |
| polyDADMAC | --- | 28000 | 0.73 | 0.96 | 15.9 | 18.2 | 13.6 | 1.6 | --- |
| polyDADMAC | --- | 165000 | 0.68 | 0.91 | 15.2 | 17.7 | 12.9 | 1.1 | --- |
| polyDADMAC | --- | 941000 | 0.73 | 0.96 | 16.2 | 19.1 | 11.1 | 1.5 | --- |

**Table S 3.** Summary of the effects of polymeric additives on the titration curves of C-S-H at various conditions. For the calculations of the Ca$^{2+}$ binding capability of negatively charged polymers at pH 12 only charged monomers are taken into account, e.g. only PAA for PVP-co-PAA.

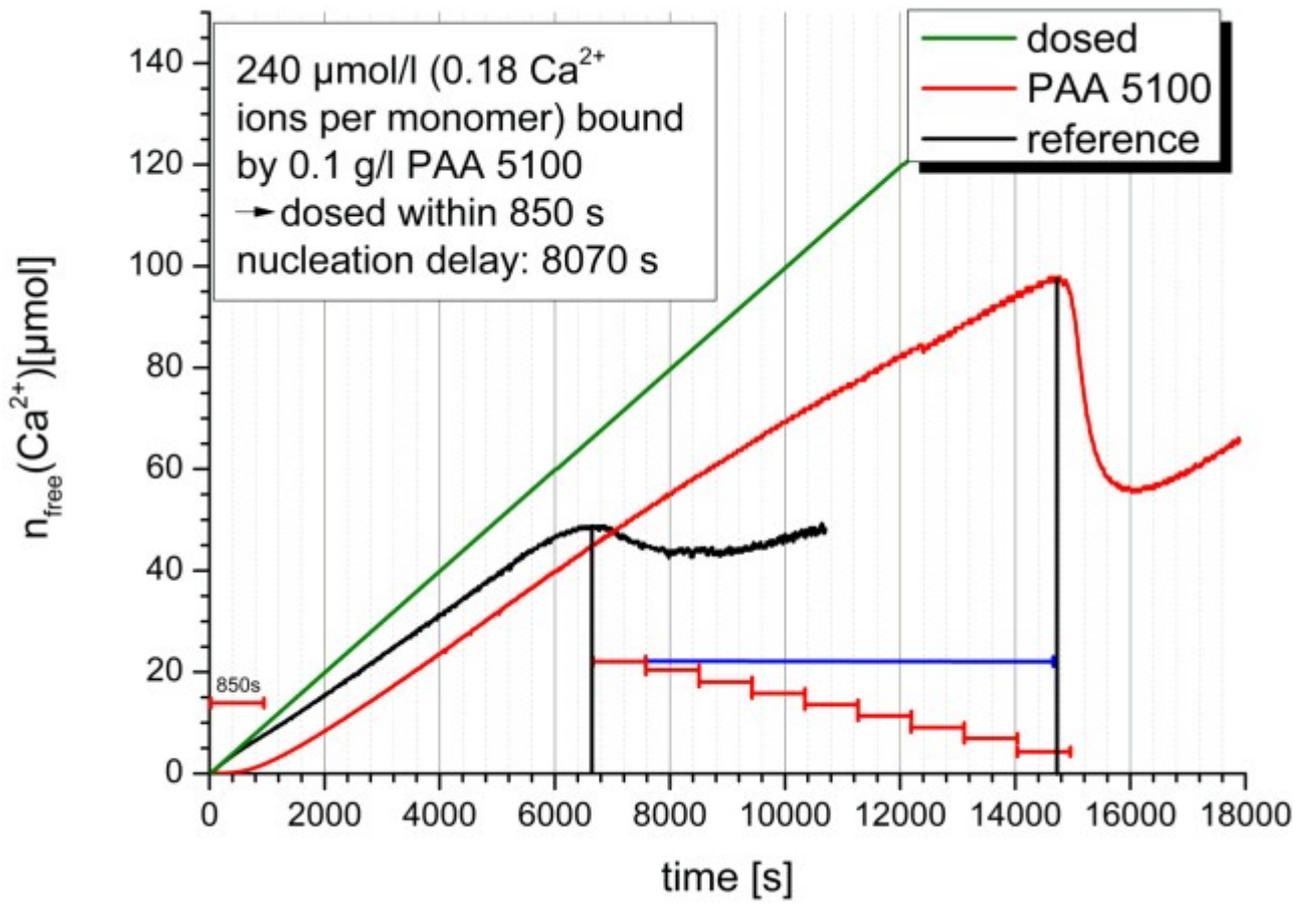

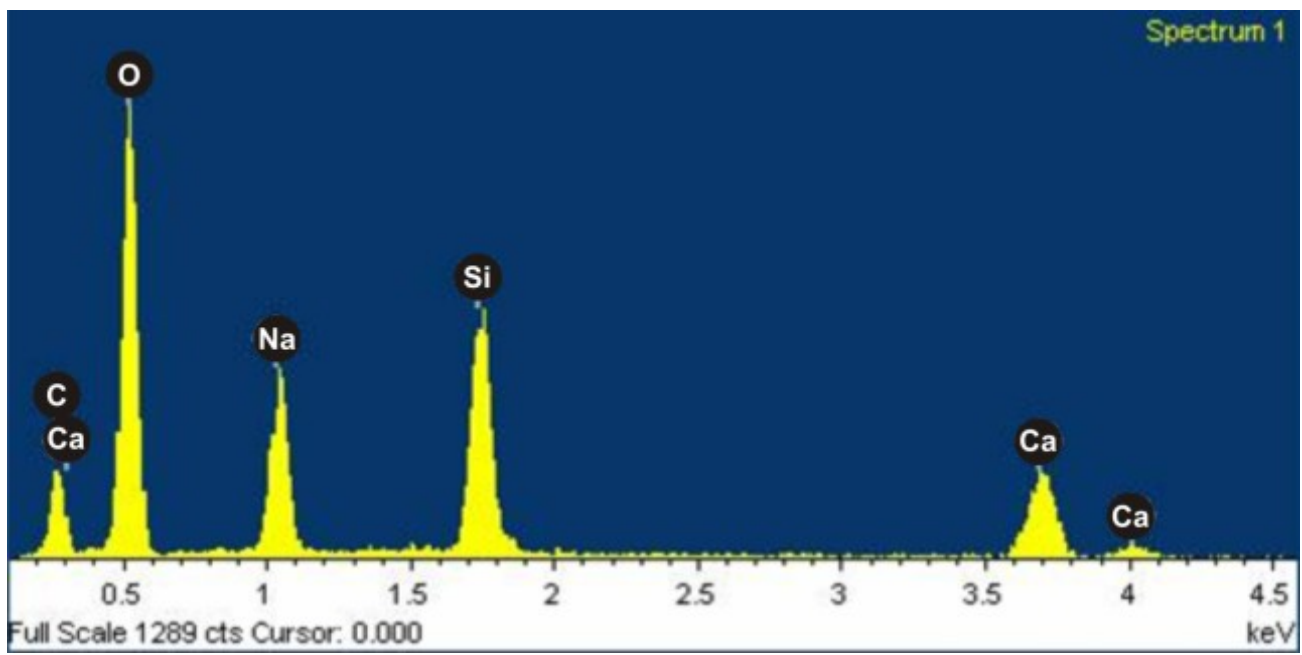

**Figure 51:** EDX analysis of the prenucleation stage, corresponding to point 1 in the inlet in Figure 6.